\documentclass[numsec,webpdf,modern,medium]{oup-authoring-template}
\onecolumn % for one column layouts

\graphicspath{{Fig/}}

\theoremstyle{thmstyleone}%
\newtheorem{theorem}{Theorem}%  meant for continuous numbers
\newtheorem{conj}{Conjecture} %%AD

\theoremstyle{thmstyletwo}%
\theoremstyle{thmstylethree}%

\usepackage{newpxtext}
\usepackage[varbb]{newpxmath}
\usepackage{titlesec}
\titleformat*{\section}{\raggedright\sffamilyfontbold\fontsize{10}{12}\selectfont}
\titleformat*{\subsection}{\raggedright\sffamilyfont\fontsize{9}{12}\selectfont}
\titleformat*{\subsubsection}{\sffamilyfont\itshape}

\usepackage{bm}
\usepackage{ragged2e} %specify table column widths
\usepackage{comment}
\usepackage{blkarray} %labelling matrix rows/cols
\usepackage{mathtools} %to read in and format matrices 
\usepackage{pdflscape} %landscape pages

\begin{document}

% Title page -------------------------------------------
%\journaltitle{Journal of the Royal Statistical Society Series A}
\journaltitle{Journal}
\DOI{DOI}
\copyrightyear{2025}
\pubyear{2025}
\access{Day Month Year}
\appnotes{Paper}

\firstpage{1}

%\subtitle{Subject Section}

\title[The bipartite structure of treatment-trial networks]{The bipartite structure of treatment-trial networks reveals the flow of information in network meta-analysis}

\author[1,$\ast$]{Annabel Davies\ORCID{0000-0003-2320-7701}}
\address[1]{\orgdiv{Bristol Medical School}, \orgname{University of Bristol}, \orgaddress{\street{Canynge Hall, 39 Whatley Road}, \postcode{BS8 2PS}, \state{Bristol}, \country{UK}}}

\authormark{Davies}

\corresp[$\ast$]{Corresponding author. \href{email:annabel.davies@bristol.ac.uk}{annabel.davies@bristol.ac.uk}}

%\received{Date}{0}{Year}
%\revised{Date}{0}{Year}
%\accepted{Date}{0}{Year}

% Abstract and keywords --------------------------------
\abstract{
Network meta-analysis (NMA) combines evidence from multiple trials comparing treatment options for the same condition. 
The method derives its name from a graphical representation of the data where nodes are treatments, and edges represent comparisons between treatments in trials. 
However, edges in this graph are limited to pairwise comparisons and fail to represent trials that compare more than two treatments.
In this paper, we describe NMA as a bipartite graph where trials define a second type of node. 
Edges then correspond to the arms of trials, connecting each trial node to the treatment nodes it compares. 
We consider an NMA model parameterized in terms of the observations in each arm. 
By linking the hat matrix of this model to the bipartite framework, we reveal how evidence flows through the arms of trials. 
We then define a random walk on the bipartite graph and propose two conjectures that relate the movement of this walker to evidence flow.
We illustrate our methods on a network of treatments for plaque psoriasis and verify our conjectures in simulations on randomly generated graphs. 
The bipartite framework provides new insights into the evidence structure of NMA and the role of individual trials in producing NMA estimates.
}

\maketitle

\section{Introduction}
Network meta-analysis (NMA) is a statistical method widely used in medical research to synthesize evidence from multiple trials that compare treatment options for the same condition (\cite{Hig:White:1996, Lu:Ades:2004, TSD2}). 
By combining direct and indirect evidence from all relevant trials, NMA produces a coherent ranking of the treatments that allows every intervention to be compared with every other. 
The results of these analyses are used routinely by policy makers to inform treatment recommendations for clinical practice (\cite{DIAS:2018}).

The use of indirect evidence in NMA facilitates comparisons between treatments that have never been directly compared in a trial. 
For example, consider two treatments $a$ and $b$ that have each been compared with a common third treatment $c$ but have not been directly compared with each other. 
Indirect evidence refers to the idea that we can infer the relationship between $a$ and $b$ via their relations to the common comparator $c$. 
That is, if we know that $a$ is more effective than $c$ (from trials comparing $a$ to $c$) and that $c$ is more effective than $b$ (from trials comparing $b$ to $c$), then we can infer that $a$ is more effective than $b$, and by how much. 

\begin{figure}
    \centering
    \includegraphics[width=0.7\linewidth]{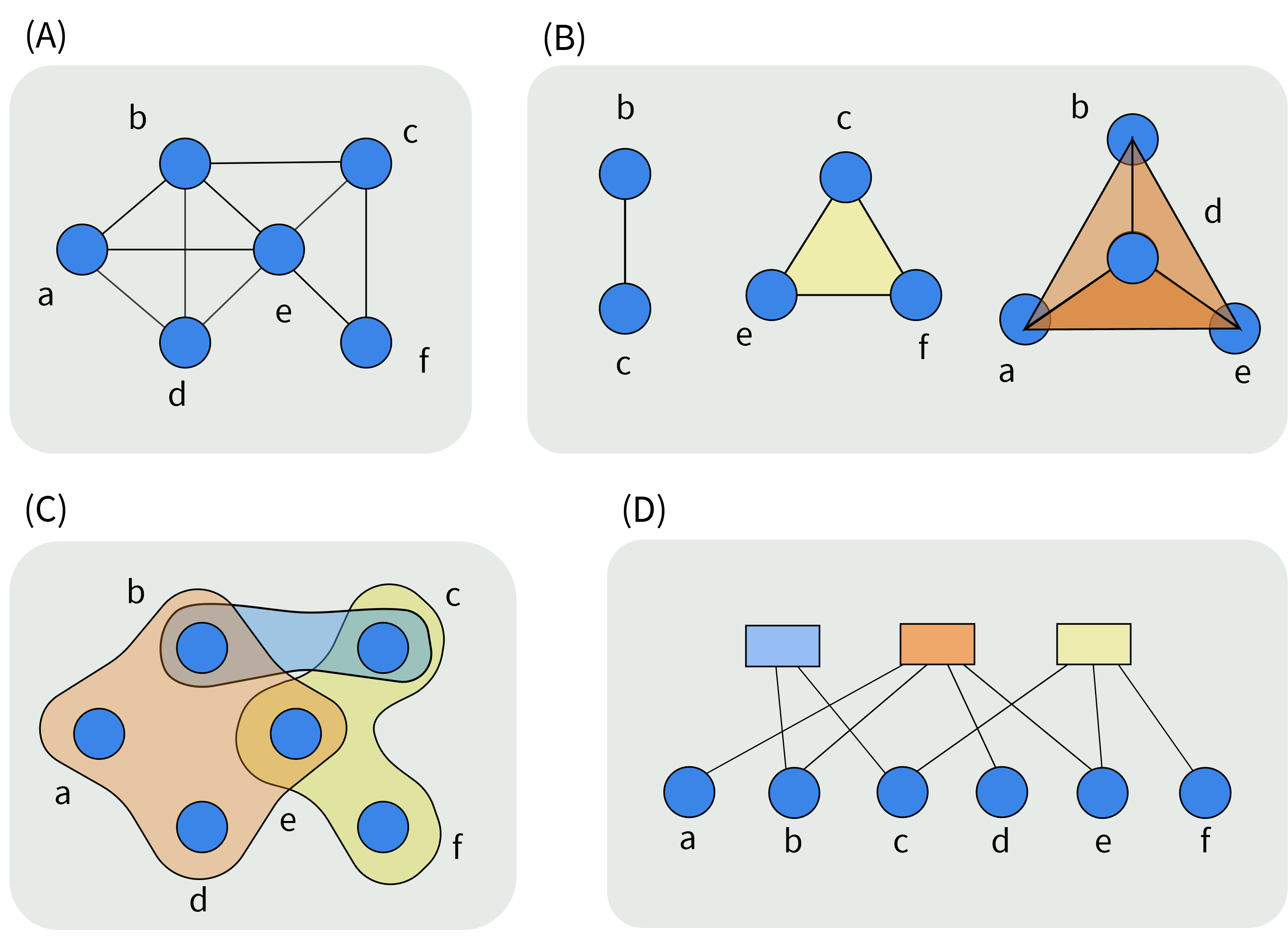}
    \caption{Four representations of an NMA involving six treatments $[a,b,c,d,e,f]$ compared in three trials; a two-arm trial $[b,c]$, a three arm trial $[c,e,f]$, and a four-arm trial $[a,b,d,e]$. Panel (A) is a graph where treatments are nodes and edges represent pairwise connections between nodes. Each trial introduces a clique (fully connected subgraph) between the treatments it compares. Panel (B) shows three simplicial complexes (with dimensions 1, 2 and 3) representing the three trials. Panel (C) is a hypergraph where hyper-edges represent trials connecting two or more nodes at once. Finally, panel (D) shows a bipartite graph where trials define a second type of node shown as rectangles. Edges represent the arms of a trial, connecting each trial node to the treatment nodes it compares. }
    \label{fig:HOI-rep}
\end{figure}

The evidence structure of an NMA is typically represented as a graph (or network), 
where nodes are the different treatment options, and the connecting edges represent comparisons between treatments in trials (\cite{Lumley:2002}). 
The set of treatments compared in a trial are referred to as the arms of the trial. 
For example, consider a set of six treatments $[a,b,c,d,e,f]$ compared in three trials; a two-arm trial comparing treatments $b$ and $c$, a three-arm trial comparing $c$, $e$, and $f$, and a four-arm trial comparing $a$, $b$, $d$, and $e$. 
Figure~\ref{fig:HOI-rep}~(A) shows the graphical representation of this data. 
An edge connecting two nodes indicates that these treatments have been directly compared in at least one trial. 
Indirect evidence between pairs of treatments is indicated by connections via intermediate nodes. 
For example, there is no trial that involves both treatments $d$ and $f$ so these nodes are not connected by an edge. 
Instead, the comparison between $d$ and $f$ is informed indirectly via intermediate comparisons to treatment $e$ shown by the two edges connecting $d$ to $e$ and $e$ to $f$.
As shown in the Figure, these treatments are also connected via longer paths of indirect evidence involving multiple intermediate nodes. 

The network structure of treatment and trials is by no means unique. 
Similar structures are found in a wide variety of systems;
the human brain can be modelled as a network of neurons connected by synapses (\cite{Sporns:2002}); 
networks of roads, railways, and tram lines form the infrastructure of cities across the globe (\cite{Gastner:2006, Sen:2003}); 
food webs map the interactions between difference species in an ecosystem (\cite{Krause:2003});
and people in society are connected by friendships, family relationships, and professional collaborations (\cite{Zachary:1976, Newman:2001a}). 
Network structures have long been studied in the related disciplines of graph theory (in mathematics) and complex networks (in statistical physics). 
Together with the application to various real world systems, these disciplines form the broader field of network science. 

Key topics in network science include the identification of influential nodes or edges (\cite{Bonacich:1987}), the robustness of networks to the removal of certain components (\cite{Albert:2000}), the flow of information through the network (\cite{LibenNowell:2008}), and the processes by which nodes become connected as the network is formed (\cite{ErdosRenyi:1960, BarabasiAlbert:1999}). 
Many of these topics are also relevant for NMA. 
For example, understanding how evidence from different trials combines to give overall estimates of treatment effects is a key challenge for NMA. 
Some trials may lack internal or external validity that mean they provide biased evidence to the network.
Understanding how the results of the NMA rely on these trials helps policy-makers to assess how confident they can be in the analysis so that they can make informed decisions and recommendations.
Previously, \cite{Konig:2013} defined the `flow of evidence' through the edges of the NMA graph. 
This visualises how the direct estimates associated with each edge combine to give the overall network estimates.
In subsequent work, \cite{Papakon:2018} used these flows to characterise the contributions of different direct and indirect paths of evidence, while \cite{Davies:2022} gave an interpretation for evidence flow in terms of the movement of a random walker on the graph.

Characterising the flow of information in NMA is made more challenging by the presence of trials that compare more than two treatments.
To include multi-arm trials in the NMA graph, we must first decompose them into their pairwise components.
Each $n$-armed trial then introduces a clique (fully connected subgraph) with $n(n-1)/2$ edges connecting every treatment node in that trial to every other. 
The resulting graph shows which treatments have been directly compared, but does not allow us to distinguish which comparisons arise from multi-arm trials rather than two-arm trials.
For example, from Figure~\ref{fig:HOI-rep}~(A) it is impossible to determine whether the edge $[a,e]$ originated from a 2-arm trial between $a$ and $e$, a 3-arm trial $[a,b,e]$ or $[a,d,e]$, or the (true) 4-arm trial $[a,b,d,e]$.
Because we cannot identify individual trials from the edges of the graph, it is not clear how the flow of information through the graph is related to the influence of these trials.

In the language of network science, edges in a graph represent `interactions' between pairs of nodes.
More specifically, they represent `first order' interactions.
The order of an interaction refers to the number of nodes it involves; an interaction between $n$ nodes has order $n-1$.
Orders greater than one (involving more than two nodes) are described as `higher-order'.
Therefore, multi-arm trials in NMA correspond to higher-order interactions (HOI) in network science; a three-arm trial is a second-order interaction, a four-arm trial is a third-order interaction, and so-on.
To accurately describe these trials, we require a higher-order representation that encodes connections between more than two nodes at once.

Panels (B)-(D) in Figure~\ref{fig:HOI-rep} show three HOI representations of our example NMA. 
In panel (B) each trial is shown as a simplex, a mathematical structure, similar to the concept of a clique, that involves a collection of fully connected nodes. 
The dimension of a simplex is equivalent to the order of an interaction; the two-arm trial $[b,c]$ is represented by a 1-simplex shown as a line connecting the two nodes, the three arm trial $[c,e,f]$ is a 2-simplex or triangle, and the four-arm trial $[a,b,d,e]$ is a 3-dimensional tetrahedron. 
A simplicial complex is a collection of simplices involving a set of common nodes that satisfies the condition of `downward closure'.
This condition requires that if an interaction exists between $n$ nodes, then every possible sub-interaction between these nodes must also exist.
For NMA, this means that if, for example, a network includes the three-arm trial $[c,e,f]$ then it must also include the two arm trials $[c,e]$, $[c,f]$, and $[e,f]$, and single arm trials of $c$, $e$, and $f$ alone.
Since this condition is rarely (if ever) met in practice, simplicial complexes are not a suitable representation of treatment-trial networks.

Another, more flexible representation of HOIs is a hypergraph, shown in panel (C) of Figure \ref{fig:HOI-rep}.
A hypergraph is a generalisation of a graph where edges can connect any number of nodes at once. 
In the context of NMA, an $n-$armed trial corresponds to a hyperedge connecting $n$ nodes. 
From Figure \ref{fig:HOI-rep} (C), it is then immediately clear that the connection between treatments $a$ and $e$, for example, originates from a 4-arm trial also involving treatments $b$ and $d$. 

Alternatively, hypergraphs can be represented as bipartite structures where the hyperedges (interactions) define a second type of node (\cite{Walsh:1975}). 
Panel (D) in Figure \ref{fig:HOI-rep} shows the bipartite graph of our example NMA.
Here, trials correspond to `top nodes' shown as rectangles and plotted above the circular treatment nodes, now referred to as `bottom nodes'. 
In this representation, edges correspond to the arms of a trial, connecting each trial node to the treatments it compares.
Many systems studied in network science have a naturally bipartite structure with the original nodes depicted as bottom nodes, and interactions represented as top nodes (\cite{Guillaume:2006}).
For example, \cite{Newman:2001a, Newman:2001b} studied a collaboration network of scientists who are connected if they are co-authors on an academic paper.
This naturally maps to a bipartite structure with scientists as the bottom nodes and papers as top nodes. 
Scientist nodes are then connected to the papers they authored. 
Similar structures include networks of actors (bottom nodes) and the films (top nodes) in which they co-star (\cite{Watts:1998}), or co-occurrence networks of words (bottom nodes) connected by the sentences (top nodes) in which they appear (\cite{Sole:2001}). 

The bipartite structure of treatment-trial networks and its potential insights have been noted recently (\cite{Davies:2022b}, \cite{Lumley:2024}), but have yet to be explored in practice. 
In this article, we formalise the bipartite framework for NMA and use this to examine the flow of information through the network. 
This goes beyond previous work on this topic by revealing how evidence flows through the arms of trials.

The paper is structured as follows: In Section \ref{sec:eg} we introduce a motivating NMA dataset of treatments for plaque psoriasis and show its representation both as a graph and a bipartite graph. 
Section \ref{sec:NMAmodels} then describes three equivalent formulations of the frequentist NMA model: the standard 'trial-level' model written in terms of the relative effects in each trial, an 'arm-level' model based on the observations in each arm, and an `aggregate' model that performs the inference in two steps.
Each model is expressed as a linear equation with a hat matrix containing the corresponding linear regression coefficients. 
The mathematical notation describing NMA as a graph, hypergraph and bipartite graph is set out in Section \ref{sec:HOI}. 
In Section \ref{sec:flow}, we demonstrate how linking the hat matrices of the different models to the different graphical representations of NMA reveals the flow of information through the network.
In particular, we link the arm-level hat matrix to the bipartite graph and use this connection to visualize evidence flow through each trial arm. 
Finally, we construct a random walk on the bipartite graph and propose two conjectures that relate the walker’s movement to evidence flow. 
In Section \ref{sec:application} we apply our approach to the motivating example, and in Section \ref{sec:simulations} we verify our conjectures on simulated datasets.
Finally, we summarize and discuss our work in Section~\ref{sec:discuss}.

\section{Motivating dataset}\label{sec:eg}

We consider a dataset from \cite{Warren:2018} comparing seven treatments for plaque psoriasis.
The network includes nine trials; three are four-arm trials, four are three-arm, and two have just two arms. 
The outcome of interest we focus on is an improvement of 75\% on the Psoriasis Area and Severity Index (PASI) scale at 12 weeks compared to baseline. 
We measure the treatment effect in each arm as the log odds of achieving this outcome. 
The relative treatment effect is a log odds ratio. The dataset is available to download from the multinma package in R (\cite{multinma}). 

\begin{figure}[b]
    \centering
    \includegraphics[width=1\linewidth]{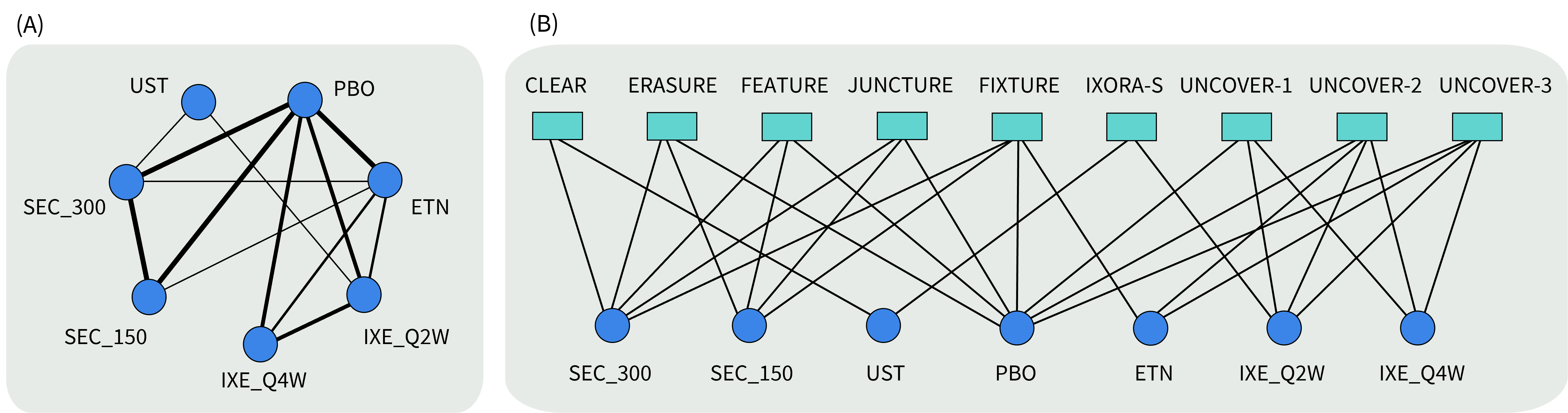}
    \caption{(A) A graph representing the plaque psoriasis NMA. Each node represents a treatment, and two nodes are connected by an edge if they have been compared in at least one trial. The thickness of each edge is proportional to the number of trials comparing those two treatments. Treatments are labelled as follows: PBO, placebo; IXE, ixekizumab; SEC, secukinumab; ETN, etanercept; UST, ustekinumab. IXE and SEC were each investigated with 2 different dosing regimens. (B) A bipartite graph representing the same dataset. Bottom nodes (circles) represent the same set of treatments as in (A) and top nodes (squares) represent trials. Trial nodes are connected to the treatment nodes they compare. Trial labels correspond to the names assigned in \cite{Warren:2018}. }
    \label{fig:psoriasis}
\end{figure}

The graph corresponding to this data is shown in Figure \ref{fig:psoriasis} (A) and comprises $7$ nodes representing the treatments and $13$ edges representing pairwise comparisons.
The thickness of each edge is proportional to the number of trials that have made that comparison.
From this figure, we can see which treatments have been directly compared but we cannot tell which comparisons come from multi-arm trials.
In contrast, the bipartite graph in Figure \ref{fig:psoriasis} (B) shows precisely which treatments are compared in each trial.
This graph comprises $16$ nodes; 9 top nodes representing the trials, and 7 bottom nodes representing the treatments. 
There are $28$ edges representing the arms of the trials.  
The number of edges connected to a node is referred to as its `degree'. 
The degree of a trial node corresponds to the number arms in that trial, whereas the degree of the treatment node tells us the number of trials in which that treatment appeared. 
For example, the degree of the JUNCTURE trial node is three, representing the three treatments in that trial; two doses of secukinumab (SEC\_150 and SEC\_300) and a placebo (PBO). 
The degree of the UST treatment node is two, meaning this treatment only appears in two trials (CLEAR and IXORA-S).

Performing an NMA on this dataset reveals the relative effects between each pair of treatments in the network.
Of particular clinical interest is the comparison between the licensed doses of interleukin-17 blockers, secukinumab 300mg (SEC\_300) and ixekizumab every 2 weeks (IXE\_Q2W). 
To make an informed decision between these treatments, policy makers need to know which trials form the main evidence base for the comparison.
A key step to achieving this is understanding how information or `evidence' flows through the network.

\section{NMA models}\label{sec:NMAmodels}

In this section, we introduce three parameterisations of the frequentist inverse-variance NMA model. 
Each parameterisation is expressed as a linear model that regresses the trial observations on a set of consistent treatment effects.
The influence of observations from the different trials is related to the structure of the network and the (inverse-variance) weight assigned to each observation in the model. 
This information is captured in the model's hat matrix which contains the linear regression coefficients.

\subsection{Set-up and notation}\label{sec:setup}

We consider a network of $N$ treatments, labelled $j=1, \hdots, N$, that are compared in $M$ trials, $i=1,\hdots, M$. Each trial $i$ compares a subset of $n_i$ treatments, $\mathcal{T}_i=[t_{i\ell}; \ell=1,\hdots n_i]\subset [1,\hdots,N]$, each associated with an observed mean $\mu_{i,t_{i\ell}}$ and variance $\sigma_{i,t_{i\ell}}^2$. 
We define relative treatment effects, or `contrasts', with reference to some trial-specific baseline treatment $t_{i1}$. 
Each trial is then associated with $n_i-1$ contrast-level observations comparing each arm $t_{i\ell\neq1}$ to the baseline,
\begin{align}\label{eq:y-mu-i}
    y_{i,t_{i1}t_{i\ell}} = \mu_{i,t_{i\ell}} - \mu_{i,t_{i1}}.
\end{align}
Collecting the arm- and contrast-level observations into the vectors $\bm{\mu}$ and $\bm{y}$ respectively, we can write Equation (\ref{eq:y-mu-i}) as
\begin{align}\label{eq:C}
    \bm{y} = \bm{C} \bm{\mu},
\end{align}
where the matrix $\bm{C}$ maps the arm-level observations in each trial onto the contrast-level.
Each trial contributes an $(n_i-1) \times n_i$ block matrix $\bm{C}_{n_i}$ whose columns represent the arms of the trial and rows represent comparisons to the baseline $t_{i1}$. 
In each row, the element corresponding to $t_{i1}$ is equal to $-1$ and the element corresponding to $t_{i\ell}$ is $+1$.
In other words, $\bm{C}_{n_i}=\left[ -\bm{1}_{n_i-1} \hspace{5pt} \bm{I}_{n_i-1} \right]$ where $\bm{1}_{n_i-1}$ is an $(n_i-1)$-column vector of ones and $\bm{I}_{n_i-1}$ is the $(n_i-1) \times (n_i-1)$ identity matrix.
These blocks are arranged diagonally in $\bm{C}$, with the remaining elements equal to zero.\footnote{Formally, $\bm{C}$ is a direct sum of the trial-level blocks, $\bm{C}=\bm{C}_{n_1}\bigoplus \bm{C}_{n_2} \bigoplus \hdots \bigoplus \bm{C}_{n_M}$.}  
We show an example of this matrix in Section \ref{sec:arm-hat-flow}.

The covariance matrix describing the sampling variances and correlations of the contrast-level observations is a block diagonal, $\bm{\Sigma}=\text{diag}(\bm{\Sigma}_1,\hdots,\bm{\Sigma}_M)$. 
Each trial $i$ contributes an $(n_i-1)\times(n_i-1)$ matrix $\bm{\Sigma}_i$ with diagonal elements equal to the variance of each $y_{i,t_{i1}t_{i\ell}}$. 
Assuming the arms are independent, these elements are equal to the sum of the arm-level variances, $v_{i,t_{i1}t_{i\ell}}=\sigma_{i,t_{i1}}^2 + \sigma_{i,t_{i\ell}}^2$. 
The off-diagonal elements are the covariances between each pair of contrast-level observations, $\text{cov}(y_{i,t_{i1}t_{i\ell}}, y_{i,t_{i1}t_{i\ell'}})$, equal to the variance in the baseline arm of that trial, $\sigma_{i,t_{i1}}^2$.

\subsection{Model assumptions}

We write $\theta_{jk}$ for the mean relative treatment effect between treatments $j$ and $k$. 
We adopt the convention $\theta_{jk}=-\theta_{kj}$ and $\theta_{jj}=0$. We also make the usual assumption that for any treatments $j$, $k$ and $l$, the relative treatment effects fulfill the consistency (transitivity) relation, 
\begin{align}\label{eq:transitivity}
    \theta_{jk} = \theta_{lk} - \theta_{lj}.
\end{align}
Assuming consistency, the relative treatment effects of all $N$ treatments are fully specified by $N-1$ `basic' parameters. 
That is, if we assign treatment 1 as the `global' baseline, it is sufficient to know $\theta_{1j}$ for $j=2,\hdots,N$. 
We collect the basic parameters in the $(N-1)$-dimensional vector, $\bm{\theta}$.

Following the standard inverse-variance model, we assume the contrast-level observations $\bm{y}$ are normally distributed about their mean effects such that,
\begin{align}\label{eq:model}
    \bm{y} \sim \mathcal{N}(\bm{X}\bm{\theta}, \bm{W}^{-1}),
\end{align}
where $\bm{X}$ is the design matrix describing which comparisons are made in each trial (see Section \ref{app:design} of the Supplement), and $\bm{W}$ is the inverse-variance weight matrix. 
In the common effect (CE) model, $\bm{W}=\bm{\Sigma}^{-1}$.
In the random effects (RE) model, we allow the `true' relative treatment effects to vary between trials. We assume trial-specific effects are drawn from a multivariate normal distribution centred on the mean effects $\bm{X}\bm{\theta}$, with between-trial covariance matrix $\bm{\Omega}$. 
The RE weight matrix in Equation (\ref{eq:model}) is then $\bm{W} = (\bm{\Sigma}+\bm{\Omega})^{-1}$. 
For our purposes the form of $\bm{\Omega}$ is not important, but the usual assumption is that all relative effects share a common between-trial heterogeneity variance, $\tau^2$. 
The covariance matrix is then a block diagonal $\bm{\Omega}=\text{diag}(\bm{\Omega}_1,\hdots,\bm{\Omega}_{M})$ where $\bm{\Omega}_i$ has dimensions $(n_i-1)\times (n_i-1)$ with diagonal entries equal to $\tau^2$ and off-diagonal entries equal to $\tau^2/2$.

\subsection{Trial-level hat matrix}

The usual frequentist estimate of the basic parameters is obtained via generalised least-squares regression or, equivalently for model (\ref{eq:model}), maximum likelihood. 
The solution is the well-known Aitken estimator (\cite{Aitken:1936}),
\begin{align}\label{eq:aitken}
    \bm{\hat{\theta}} = \bm{H} \bm{y},
\end{align}
where we label $\bm{H}$ the `trial-level' hat-matrix, given by 
\begin{align}\label{eq:H-trial}
    \bm{H} = (\bm{X}' \bm{W} \bm{X})^{-1}\bm{X}'\bm{W},
\end{align}
with dimensions $(N-1) \times \sum_{i=1}^{M} (n_i-1)$. In Equation (\ref{eq:H-trial}), we use the notation $\bm{X}'$ to indicate the transpose of matrix $\bm{X}$.

Each row of $\bm{H}$ contains the coefficients of a linear equation describing how the estimates of each basic parameter, $\hat\theta_{1j}$, depend on the contrast-level observations from each trial, $y_{i,t_{i1}t_{i\ell}}$. 
To obtain the $N(N-1)/2$ relative effects between every pair of treatments, we apply the consistency equations in (\ref{eq:transitivity}) to the estimates of the basic parameters in Equation (\ref{eq:aitken}).  
Applying these equations to the right hand side of Equation (\ref{eq:aitken}) yields the ``full" hat matrix containing the coefficients for the full set of relative treatment effects.
For example, the coefficients for the relative effect $\hat\theta_{jk}$ ($j,k\neq 1$) are obtained by subtracting the row of the hat matrix corresponding to the basic comparison $1j$ from the row corresponding to $1k$,
\begin{align}\label{eq:hat-consistency}
    \bm{H}^{(jk)} = \bm{H}^{(1k)}-\bm{H}^{(1j)}.
\end{align}

\subsection{Arm-level hat matrix}

Using the relationship between the contrast-level and arm-level observations, we can instead write the model in terms of the vector of means in each arm $\bm{\mu}$. 
Substituting Equation (\ref{eq:C}) into Equation (\ref{eq:aitken}), we obtain
\begin{align}
    \bm{\hat{\theta}} = \bm{H}_{\top\bot}\bm{\mu},
\end{align}
where we define the arm-level\footnote{In the terminology set out by \cite{White:2019}, the arm-level model we describe is not `arm-based'. That is, we apply modeling assumptions to parameters representing relative effects (contrasts) rather than parameters representing arm means. Our model is simply a parameterization of the standard contrast-based model in terms of the arm-level data. This is what \cite{White:2019} calls a contrast-based model with an arm-based likelihood. } hat matrix $\bm{H}_{\top\bot}$ as a projection of the trial-level hat matrix,
\begin{align}\label{eq:H-bi}
    \bm{H}_{\top\bot} = \bm{H}\bm{C} = (\bm{X}' \bm{W} \bm{X})^{-1}\bm{X}'\bm{W}\bm{C},
\end{align}
with dimensions $(N-1)\times \sum_i n_i$. The use of the notation `$\top\bot$' will become apparent in later sections when we connect the hat matrices to the different graphical representations of NMA.

Each row of $\bm{H}_{\top\bot}$ contains the coefficients of a linear equation describing how the estimates of each basic parameter $\hat\theta_{1j}$ depend on the observations in each arm of each trial, $\mu_{i,t_{i\ell}}$. 
Again, we obtain the coefficients for every pairwise comparison by applying the consistency equations in (\ref{eq:hat-consistency}).

\subsection{Aggregate hat matrix}\label{sec:agg-hat}
Another alternative formulation of the model is to perform the regression in two steps (\cite{Lu:2011, Krahn:2013}). 
In the first step, trials comparing the same treatments are combined to obtain a set of direct estimates. 
Then, the direct estimates are entered as the observations in an NMA. 
This is referred to as an `aggregate' model. 
In particular, we use a `graph-theoretical' version of the aggregate model that accounts for correlations due to multi-arm trials before combining the direct evidence on each pairwise comparison.
We describe the model in Section \ref{app:agg} of the Supplement and refer to 
to \cite{Davies:2022} and its Supplementary Materials for further details.
The key points are summarized below.

In the first step, we account for correlations by adjusting the variances associated with each comparison in each multi-arm trial.
Using these adjusted weights, we then perform a pairwise meta-analysis (weighted mean) for each pair of treatments that have been compared in at least one trial.
The resulting direct estimates are collected in the vector $\bm{\hat\theta^{\text{dir}}}$ and their associated weights in the diagonal matrix $\bm{W}_\bot=\text{diag}(w_{jk})$.

In the second step, we perform an NMA using the direct estimates.
The network estimates of the $N-1$ basic parameters are obtained via the linear equation
\begin{align}
    \bm{\hat\theta} = \bm{H_\bot}\bm{\hat\theta^{\text{dir}}},
\end{align}
where we write $\bm{H_\bot}$ for the aggregate hat matrix, defined as
\begin{align}\label{eq:H_bot}
    \bm{H_\bot} = \bm{C}_N (\bm{B'_\bot}\bm{W_\bot}\bm{B_\bot})^{+}\bm{B'_\bot}\bm{W_\bot}.
\end{align}
The matrix $\bm{B_\bot}$ is an oriented edge-vertex incidence matrix. 
Each column corresponds to a treatment and each row corresponds to a direct estimate (edge) between two treatments.
Without loss of generality, each row representing an edge between treatments $j$ and $k$ is given an arbitrary orientation, with a $-1$ in column $j$ and $+1$ in column $k$. 
All other entries are 0.
We return to this matrix in Section \ref{sec:simple} below. 
Recall from Section \ref{sec:setup} the definition of matrix $\bm{C}_{N}=\left[ -\bm{1}_{N-1} \hspace{5pt} \bm{I}_{N-1} \right]$. 
In Equation (\ref{eq:H_bot}), this matrix ensures that $\bm{H_\bot}$ has the correct dimensions for projecting onto the $N-1$ basic parameters. 
Finally, the matrix operation `$^{+}$' indicates its Moore-Penrose pseudo-inverse.  

The aggregate hat matrix has $(N-1)$ rows representing the comparison of each treatment to the global baseline.
The number of columns is equal to the number of direct estimates (the number of edges in the NMA graph).
Each row contains the coefficients of a linear equation describing how the estimates of each basic parameter $\hat\theta_{1j}$ depend on the direct estimates associated with each edge. 
As before, we obtain the full version of the aggregate hat matrix by applying the consistency equations in (\ref{eq:hat-consistency}).

\section{Network representations}\label{sec:HOI}
In this section, we set out the mathematical descriptions of the graph, hypergraph and bipartite graph representations of treatments and trials. To keep track of how the notation and terminology in these different frameworks correspond to one another and to concepts in NMA, we provide a summary in Table \ref{tab:nma_notation} for the reader to refer back to.

\subsection{Graphs}\label{sec:simple}

Traditionally, treatments and trials in an NMA are represented as a graph, $G_\bot=[V_\bot,E_\bot]$, where $V_\bot=[v_j; j=1,\hdots,N]$ is a set of $N$ nodes representing the treatments and $E_\bot$ is a set of $K_\bot$ edges connecting pairs of nodes.
An edge $[v_j,v_k]\in E_\bot$ is defined by the two nodes it connects and is a natural representation of a 2-arm trial. 
However, to represent multi-arm trials, we must first decompose them into their pairwise components. 
Each multi-arm trial then introduces a fully connected subgraph, known as a `clique', between the treatment nodes in that trial. 
For instance, returning to our example from the Introduction, the three trials $[b,c], [c,e,f],$ and $[a,b,d,e]$ can be decomposed into the edgeset $E_\bot=[[b,c], [c,e],[c,f],[e,f],[a,b],[a,d],[a,e],[b,d],[b,e],[d,e]]$ (shown in panel (A) of Figure \ref{fig:HOI-rep}).

\begin{table}
    \centering 
    \footnotesize
    \caption{Terminology and notation in NMA and network theory. }
    \tabcolsep=8pt
    \begin{tabular}{p{0.19\textwidth}p{0.22\textwidth}p{0.23\textwidth}p{0.2\textwidth}}
    \toprule
         NMA  & Graph  & Hypergraph & Bipartite graph  \\
         \midrule
        Treatments \newline $j=1,\hdots,N$ &  Nodes \newline $V_\bot=[v_j]$ & Nodes \newline $V_\bot=[v_j]$ & Bottom nodes \newline $V_\bot=[v_j]$\\[2.5em]
        Trials \newline $i=1,\hdots,M$ &  Cliques\footnotemark[1]  & Hyperedges \newline $\mathcal{H}=[\mathcal{H}_i]$ & Top nodes \newline $V_\top=[u_i]$ \\[2.5em]
        Trial arms \newline $\mathcal{T}_i=[t_{i\ell}; \ell=1,\hdots,n_i]$ & Edges in a clique\footnotemark[1] \newline $[[v_{i1},v_{i2}],\hdots,[v_{in_i-1},v_{in_i}]]$  & Nodes connected by \newline $\mathcal{H}_i=[v_{i\ell}; \ell=1,\hdots,n_i]$ & Edges connecting $u_i$ to adj\-acent bottom nodes \\[2.5em]
        $n_i$, number of arms in trial $i$ & Size of clique & Number of nodes con\-nected by hyperedge $\mathcal{H}_i$ & Degree of top node $u_i$\\
    \botrule
    \end{tabular}
    \label{tab:nma_notation}
    \justifying{\small{
    \footnotemark[1]In a graph, edges represent comparisons between treatments in trials. An $n_i$-armed trial gives rise to a clique (fully connected subgraph) of size $n_i$. However, unlike the hypergraph and bipartite graph, there is no 1:1 mapping between a component in the graph and a trial or trial-arm.}}
\end{table}

Mathematically, a graph is described by its adjacency matrix $\bm{A}_\bot$. This is a square $N \times N$ matrix where each row and column represents a node. An element $A_{\bot jk}$ is equal to 1 if nodes $v_j$ and $v_k$ are connected by an edge, otherwise it is 0. The adjacency matrix of the graph in Figure \ref{fig:HOI-rep} (A) is 
\begin{align}
\bm{A}_\bot=
\begin{blockarray}{ccccccc}
 & a & b & c & d & e & f \\
\begin{block}{c(cccccc)}
  a & 0 & 1 & 0 & 1 & 1 & 0 \\
  b & 1 & 0 & 1 & 1 & 1 & 0 \\
  c & 0 & 1 & 0 & 0 & 1 & 1 \\
  d & 1 & 1 & 0 & 0 & 1 & 0 \\
  e & 1 & 1 & 1 & 1 & 0 & 1 \\
  f & 0 & 0 & 1 & 0 & 1 & 0 \\
\end{block}
\end{blockarray}
\hspace{2pt},
\end{align}
where we have labelled the rows and columns by the nodes they represent. 
The sum of the entries in each row and column of $\bm{A}_\bot$ is equal to the degree of that node. 
For example, the entries of row $c$ and column $c$ each sum to 3 which correpsonds to the number of edges connected to node $c$ in Figure \ref{fig:HOI-rep} (A).
As we will see later, for weighted graphs, one can define a weighted adjacency matrix where each element $A_{\bot jk}$ is equal to the weight associated with the edge $[v_j,v_k]$. 
The weighted degree of a node is then the sum of weights in the edges connected to that node.

A graph can also be described by its incidence matrix $\bm{B}_\bot$ which encodes relationships between nodes and edges. 
We focus on the $K_\bot \times N$ edge-vertex incidence matrix where each row represents an edge $\in E_\bot$  and each column represents a node $\in V_\bot$.\footnotemark[2]\footnotetext[2]{In network theory, one often comes across the $N\times K_\bot$ vertex-edge incidence matrix which is the transpose of the edge-vertex incidence matrix. 
Both matrices may be referred to as simply the `incidence matrix' so care must be taken to understand the definition being used in any given context. 
Throughout this article we will define all incidence matrices in their edge-vertex orientation.} 
The elements of $\bm{B}_\bot$ are defined such that the row representing edge $[v_j,v_k]$ contains a 1 in both columns $j$ and $k$. 
All other entries are zero. For our example, we have 
\begin{align}\label{eq:incidence}
\bm{B}_\bot=
\begin{blockarray}{ccccccc}
 & a & b & c & d & e & f \\
\begin{block}{c(cccccc)}
  \lbrack b,c\rbrack & 0 & 1 & 1 & 0 & 0 & 0\\
  \lbrack c,e\rbrack & 0 & 0 & 1 & 0 & 1 & 0\\
  \lbrack c,f\rbrack & 0 & 0 & 1 & 0 & 0 & 1\\
  \lbrack e,f\rbrack & 0 & 0 & 0 & 0 & 1 & 1\\
  \lbrack a,b\rbrack & 1 & 1 & 0 & 0 & 0 & 0\\
  \lbrack a,d\rbrack & 1 & 0 & 0 & 1 & 0 & 0\\
  \lbrack a,e\rbrack & 1 & 0 & 0 & 0 & 0 & 1\\
  \lbrack b,d\rbrack & 0 & 1 & 0 & 1 & 0 & 0\\
  \lbrack b,e\rbrack & 0 & 1 & 0 & 0 & 1 & 0\\
  \lbrack d,e\rbrack & 0 & 0 & 0 & 1 & 1 & 0\\
\end{block}
\end{blockarray}
\hspace{2pt},
\end{align}
where the sum of elements in each row is 2 and the sum of elements in each column is the degree of the node it represents. 
The matrix in \eqref{eq:incidence} is an unoriented incidence matrix describing undirected edges. 
For a directed network, we can instead define an oriented version where an edge directed from $v_j$ to $v_k$ has a $-1$ in column $j$ and a $+1$ in column $k$. Each row then sums to 0. 
It is this oriented version of the edge-vertex incidence matrix that appears in the definition of the aggregate hat matrix in Equation (\ref{eq:H_bot}).

The benefit of representing NMA as a graph is that it allows us to use the many well-established tools and techniques from graph theory. 
However, by  encoding only pairwise connections, the graph fails to accurately represent multi-arm trials.

\subsection{Hypergraphs}
Hypergraphs offer the most flexible representation of higher-order interactions and therefore, provide an accurate description of multi-arm trials in NMA. 
A hypergraph $G_\mathcal{H}=[V_\bot, \mathcal{H}]$ is a generalisation of a graph, defined by the original nodeset $V_\bot$ and a set of hyperedges $\mathcal{H}=[\mathcal{H}_i]$ that connect groups of nodes. 
Unlike an edge in a graph, which represents a pairwise connection, a hyperedge can connect any number of nodes in $V_\bot$. 
A hyperedge is defined by the subset of nodes it connects, $\mathcal{H}_i=[v_{i\ell}; \ell=1,\hdots,n_i]$, and therefore corresponds naturally to a trial comparing $n_i$ treatments.

An adjacency matrix tells us whether two nodes are connected but does not provide any information about the edge from which the connection derives. 
To fully describe a hypergraph therefore requires an incidence matrix, $\bm{B}$. Similar to the incidence matrix of the graph, columns of $\bm{B}$ represent nodes $\in V_\bot$ while rows represent hyperedges $\in \mathcal{H}$. For the row representing hyperedge $\mathcal{H}_i$, the element $B_{ij}$ is equal to 1 if node $v_j$ is connected to that hyperedge ($v_j \in \mathcal{H}_i$) and is 0 otherwise. The incidence matrix of the hypergraph in Figure \ref{fig:HOI-rep} (C) is therefore
\begin{align}\label{eq:hyper-incidence}
\bm{B}=
\begin{blockarray}{ccccccc}
 & a & b & c & d & e & f \\
\begin{block}{c(cccccc)}
  \mathcal{H}_1 & 0 & 1 & 1 & 0 & 0 & 0\\
  \mathcal{H}_2 & 0 & 0 & 1 & 0 & 1 & 1\\
  \mathcal{H}_3 & 1 & 1 & 0 & 1 & 1 & 0\\
\end{block}
\end{blockarray}
\hspace{2pt},
\end{align}
where $\mathcal{H}_1=[b,c]$, $\mathcal{H}_2=[c,e,f]$, and $\mathcal{H}_3=[a,b,d,e]$.
Here, the elements in each row sum to the number of nodes connected by that hyperedge and each column sums to the number of hyperedges connected to that node (known as the hyper-degree of the node).

Although many concepts and results from graph theory can be extended to hypergraphs, this often comes with increased mathematical complexity.
For NMA, this is compounded by the fact that we often have multiple trials that have compared the same set of treatments. 
This requires a so-called `multi-hypergraph' which allows multiple hyperedges to connect the same set of nodes.
This is known as `edge-multiplicity' and further increases the  complexity of these structures.

\subsection{Bipartite graphs}\label{sec:bipartite}

A bipartite graph $G_{\top\bot}=[V_\top, V_\bot, E_{\top\bot}]$ contains two disjoint sets of nodes, $V_\top$ and $V_\bot$, referred to as the top and bottom nodes respectively.  
$E_{\top\bot} \subseteq V_\top \times V_\bot$ defines a set of edges connecting pairs of nodes of opposite types. 
The structure is such that edges only exist between top and bottom nodes; there can be no edge between two top nodes or between two bottom nodes. 
The terminology derives from the standard visualization of bipartite graphs where the two sets of nodes are arranged in two horizontal lines, with top nodes appearing above bottom nodes.

Hypergraphs can be represented as bipartite structures where bottom nodes are the original node set $V_\bot$, top nodes correspond to the hyperedges  $V_\top=\mathcal{H}$, and edges capture pairwise connections between nodes and hyperedges. 
We retain the notation $v_j \hspace{5pt} (j=1,\hdots,N)$ to label bottom nodes, and write $u_i \hspace{5pt} (i=1,\hdots,M)$ for top nodes such that node $u_i$ corresponds to the hyperedge $\mathcal{H}_i$. 
We write $[u_i, v_j]$ for the edge connecting top node $u_i$ to bottom node $v_j$. 
In NMA, the bottom nodes are treatments and top nodes are trials. 
Edges then represent the arms of a trial, connecting trial nodes to the treatment nodes they compare. 
The degree of a trial node is equal to the number of arms in that trial, while the degree of a treatment node tells us the number of trials that treatment is involved in. 

By representing interactions themselves as nodes, the bipartite framework re-establishes edges as encoding pairwise connections. 
This means that we can describe the bipartite graph using an adjacency matrix $\bm{A}$ that tells us whether or not two nodes are connected. 
Given $M$ top nodes and $N$ bottom nodes, the adjacency matrix of the bipartite graph has dimensions $(M+N)\times(M+N)$. 
The first $M$ rows and columns represent the top nodes (trials) and the final $N$ rows and columns represent the bottom nodes (treatments). 
As before, an element of $\bm{A}$ is equal to 1 if the nodes corresponding to that row and column are connected, and is 0 otherwise. 
For the example in Figure \ref{fig:HOI-rep} (D), we find
\begin{align}\label{eq:bi-adj}
\bm{A}=
\begin{blockarray}{cccccccccc}
 & u_1 & u_2 & u_3 & \phantom{,}a\phantom{,} & \phantom{,}b\phantom{,} & \phantom{,}c\phantom{,} & \phantom{,}d\phantom{,} & \phantom{,}e\phantom{,} & \phantom{,}f\phantom{,} \\
\begin{block}{c(ccccccccc)}
  u_1 & 0 & 0 & 0 & 0 & 1 & 1 & 0 & 0 & 0 \\
  u_2 & 0 & 0 & 0 & 0 & 0 & 1 & 0 & 1 & 1\\
  u_3 & 0 & 0 & 0 & 1 & 1 & 0 & 1 & 1 & 0\\
  a & 0 & 0 & 1 & 0 & 0 & 0 & 0 & 0 & 0\\
  b & 1 & 0 & 1 & 0 & 0 & 0 & 0 & 0 & 0\\
  c & 1 & 1 & 0 & 0 & 0 & 0 & 0 & 0 & 0\\
  d & 0 & 0 & 1 & 0 & 0 & 0 & 0 & 0 & 0\\
  e & 0 & 1 & 1 & 0 & 0 & 0 & 0 & 0 & 0\\
  f & 0 & 1 & 0 & 0 & 0 & 0 & 0 & 0 & 0\\
\end{block}
\end{blockarray}
\hspace{2pt},
\end{align}
where $u_1=[b,c]$, $u_2=[c,e,f]$, and $u_3=[a,b,d,e]$.
Equation \eqref{eq:bi-adj} reveals a block diagonal structure that is characteristic of the adjacency matrices of bipartite graphs.
Because  two nodes of the same type cannot be connected, the $M\times M$ and $N \times N$ diagonal blocks representing connections between pairs of top nodes and between pairs of bottom nodes are equal to zero. 
On inspection, we observe that the $M\times N$ upper right block is equal to the incidence matrix of the hypergraph in Equation \eqref{eq:hyper-incidence}. 
The $N\times M$ lower left block is equal to its transpose, $\bm{B}'$. 
The adjacency matrix therefore takes the form
\begin{align}\label{eq:AB}
    \bm{A} = \begin{pmatrix}
        0 & \bm{B} \\
        \bm{B}' & 0
    \end{pmatrix}.
\end{align}
In the bipartite framework, the matrix $\bm{B}$ is referred to as the `biadjacency matrix' and encodes all the information about the graph. 
Therefore, it is often the biadjacency matrix rather than the full adjacency matrix $\bm{A}$ that is used to describe bipartite graphs. 

It can also be useful to define the edge-vertex incidence matrix of the bipartite graph. 
To distinguish this from the biadjacency matrix, we label it $\bm{B}_{\top\bot}$. 
It is defined in the same way as the incidence matrix of a graph, with rows representing edges $\in E_{\top\bot}$ (trial arms), and columns representing nodes. 
The first $M$ columns correspond to top nodes $\in V_{\top}$ (trials) and the last $N$ columns correspond to bottom nodes $\in V_\bot$ (treatments). 
For the row representing edge $[u_i, v_j]$ there is a $1$ in both columns $i$ and $i+j$. 
All other entries are 0. As before, an oriented version of the incidence matrix can be defined by setting the element in column $i$ equal to $-1$. 
In Section \ref{app:bi-incidence} of the Supplement, we show $\bm{B}_{\top\bot}$ for the example in Figure \ref{fig:HOI-rep} (D). 
We summarize our notation in Table \ref{tab:graph_notation}.

\begin{table}
    \centering
    \caption{A summary of notation for the three different types of graph.}
    \begin{tabular}{p{0.19\textwidth} p{0.19\textwidth} p{0.19\textwidth} p{0.19\textwidth}}
    \toprule
         & Graph & Hypergraph & Bipartite graph  \\
         \midrule 
        Graph & $G_\bot$ & $G_{\mathcal{H}}$ & $G_{\top\bot}$\\[0.5em]
        Nodeset & $V_\bot$ & $V_\bot$ & $[V_\top, V_\bot]$\\[0.5em]
        Edgeset & $E_\bot \subseteq V_\bot \times V_\bot$ & $\mathcal{H}$ & $E_{\top\bot} \subseteq V_\top \times V_\bot$\\[0.5em]
        Node & $v_j \in V_\bot$ & $v_j \in V_\bot$ & $u_i \in V_\top$, $v_j \in V_\bot$\\[0.5em]
        Edge & $[v_j, v_k]$ & $[v_j, v_k, v_l, \hdots]$ & $[u_i,v_j]$ \\[0.5em]
        Number of nodes & $N$ & $N$ & $M+N$\\[0.5em]
        Number of edges & $K_\bot$ & $M$ & $K_{\top\bot}$\\
        Adjacency matrix & $\bm{A}_\bot$ & - & $\bm{A}=\begin{pmatrix}
            0 & \bm{B} \\
            \bm{B}'& 0
        \end{pmatrix}$\\
        Incidence matrix & $\bm{B}_\bot$ & $\bm{B}$ & $\bm{B}_{\top\bot}$\\%[0.5em]
        %Hat matrix (flow of evidence) & $\bm{H}_{\bot}$ & - & $\bm{H}_{\top\bot}$\\
    \botrule
    \end{tabular}
    \label{tab:graph_notation}
\end{table}

The relation in Equation \eqref{eq:AB} reveals the 1:1 mapping between hypergraphs and bipartite graphs. 
Despite this equivalence, the two representations have different mathematical properties. 
By encoding only pairwise connections between nodes, bipartite graphs more easily inherit operations and results from standard graph theory. Moreover, the bipartite framework naturally facilitates hyperedge multiplicity; multiple top nodes may connect the same set of bottom nodes without introducing any additional complexity. For these reasons, we select bipartite graphs as the most useful representation of higher-order structures in NMA networks. 

\subsubsection{Unipartite projection}
The unipartite graph $G_\bot=(V_\bot, E_\bot)$ associated with the bipartite graph $G_{\top\bot}=(V_\top, V_\bot, E_{\top\bot})$ is called the unipartite-projection of $G_{\top\bot}$. 
Throughout the remainder of the article, we will use the term `unipartite graph' to distinguish it from the bipartite graph.
Other terminology includes the one-mode-, clique-based, bottom-, or $\bot-$projection (\cite{Guillaume:2004}). 
In this projection, edge $[v_j,v_k]$ is in $E_\bot$ if bottom nodes $v_j$ and $v_k$ are both connected to the same top node in $G_{\top\bot}$. 
Therefore, each top node in $G_{\top\bot}$ induces a clique in $G_\bot$ between the bottom nodes to which it is linked. 
For example, in Figure \ref{fig:HOI-rep}, the graph in panel (A) is the unipartite-projection of the bipartite graph in panel (D). 
The central top node in panel (D) is connected to four bottom nodes $a,b,d,e$.
In panel (A) this corresponds to the fully connected subgraph linking each pair of these nodes to each other.

Every bipartite graph corresponds to a unique unipartite-projection. 
However, the reverse does not hold: a single unipartite projection can result from multiple distinct bipartite graphs.
In general, therefore, it is not possible to reconstruct the original bipartite graph from a given unipartite graph. 
Indeed, this is the problem we described earlier that means we cannot identify multi-arm trials from the unipartite representation of NMA.

\section{Flow of information}\label{sec:flow}

In this section we explore the flow of information through the network by linking the different hat matrices in Section \ref{sec:NMAmodels} to the different network representations in Section \ref{sec:HOI}. 
We begin with a review of previous work connecting the aggregate hat matrix to evidence flow on the unipartite graph. 
We present a theorem relating this evidence flow to the movement of a random walker on the graph.
Building on these previous results, we link the arm-level hat matrix to the bipartite graph and use this connection to visualize evidence flow through each trial arm.
Finally, we construct a higher order random walk on the bipartite graph and propose two conjectures that relate the walker's movement to evidence flow across both the bipartite and unipartite graphs.

\subsection{Unipartite evidence flow}

\subsubsection{The aggregate hat matrix}

The aggregate hat matrix $\bm{H}_\bot$ contains the coefficients of a linear equation relating the treatment effect estimates to the direct estimates.
Each direct estimate is associated with an edge in the unipartite NMA graph.
\cite{Konig:2013} showed that each row of $\bm{H}_\bot$ representing a particular treatment comparison $jk$ describes a directed network from node $v_j$ to node $v_k$ on this graph.
Each element of the row corresponds to an edge $[v_l, v_m] \in E_\bot$. 
The magnitude of flow between nodes $v_l$ and $v_m$ is given by the absolute value of this element and the direction of flow is determined by its sign. 
We refer to \cite{Konig:2013, Davies:2022}, and \cite{Rucker:2024} for a discussion of the properties of these flow networks.

As an illustration, we consider a fictional example  with $M=5$ trials labelled $[1,2,3,4,5]$, and $N=4$ treatments labelled $[a,b,c,d]$. 
The network contains three two-armed trials ($1,3,$ and $5$) and two three-armed trials ($2$ and $4$). 
Table \ref{Tab:EG} shows the arm-level data associated with each trial. 

\begin{table}[h]
\centering
\caption{Arm level data for our fictional example network.\label{Tab:EG}}
\begin{tabular}{llll}
\toprule
Trial, $i$ & Arms, $\mathcal{T}_i=[t_{i\ell}]$ & Means, $[\mu_{i,t_{i\ell}}]$ & Variances, $[\sigma_{i,t_{i\ell}}^2]$\\
\midrule
1 & $[a, b]$ & [1.0, 2.0] & [0.2, 0.3]\\
2 & $[a, b, c]$ & [0.8, 2.1, 1.5] & [0.4, 0.6, 0.5]\\
3 & $[a, d]$ & [1.2, 1.5] & [0.8, 0.9]\\
4 & $[b, c, d]$ & [2.0, 1.6, 1.2] & [1.2, 1.1, 1.0]\\
5 & $[c, d]$ & [1.4, 0.9] & [0.5, 0.4]\\
\botrule
\end{tabular}
\end{table}

\begin{figure}
    \centering
    \includegraphics[width=0.6\linewidth]{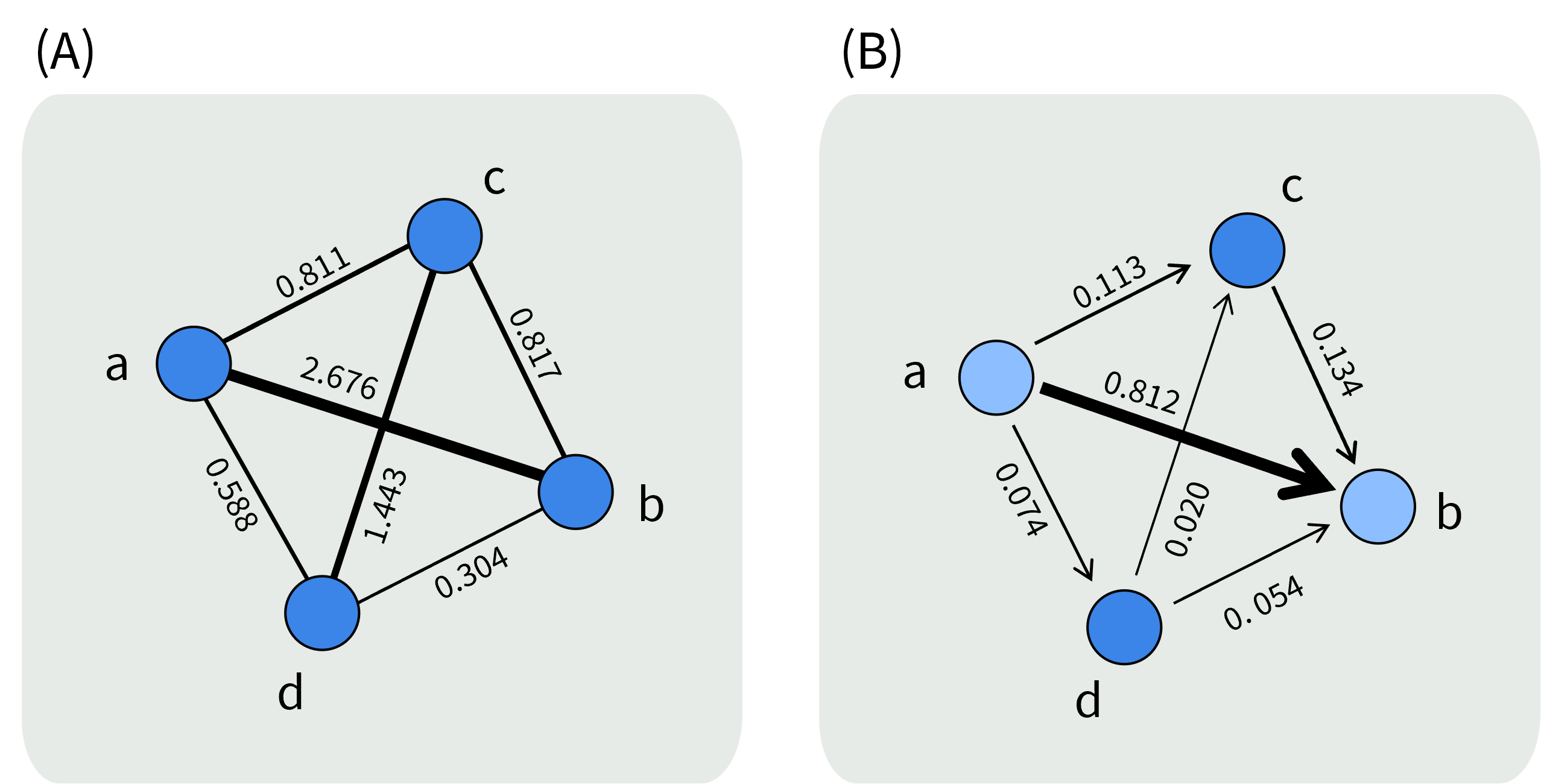}
    \caption{(A) Weighted unipartite graph for the fictional example network described in Table \ref{Tab:EG}. The thickness of each edge $[v_j,v_k]$ is proportional to its aggregate weight $w_{jk}$ in the corresponding element of $\bm{W_\bot}=\text{diag}(w_{jk})$. These weights are shown in the edge labels. (B) The evidence flow network for the comparison of treatments $a$ and $b$. The thickness of each edge is proportional to the corresponding element of the aggregate hat matrix $\bm{H}_\bot$ shown in the edge labels. The direction of each edge is given by the sign of that element.}
    \label{fig:EG-uni}
\end{figure}

We assume a CE model ($\tau=0$), and choose treatment $a$ as the global baseline. 
Panel (A) of Figure \ref{fig:EG-uni} shows the unipartite graph of this example. 
The graph has $K_\bot=6$ edges, $E_\bot = [[a,b],[a,c], [a,d],\allowbreak [b,c], [b,d], [c,d]]$, with weights $\bm{W}_\bot =\text{diag}(2.676, 0.811, 0.588, 0.817, 0.304, 1.443)$.
Following the convention set out in Section \ref{sec:agg-hat}, the oriented incidence matrix of this graph is 
\begin{align}
    \bm{B_\bot} = \begin{pmatrix}
        % latex table generated in R 4.4.1 by xtable 1.8-4 package
% Mon May 19 17:20:13 2025
 \mathmakebox[\widthof{$-1$}][c]{-1} & \mathmakebox[\widthof{$-1$}][c]{\phantom{-}1} & \mathmakebox[\widthof{$-1$}][c]{\phantom{-}0} & \mathmakebox[\widthof{$-1$}][c]{\phantom{-}0} \\ 
  \mathmakebox[\widthof{$-1$}][c]{-1} & \mathmakebox[\widthof{$-1$}][c]{\phantom{-}0} & \mathmakebox[\widthof{$-1$}][c]{\phantom{-}1} & \mathmakebox[\widthof{$-1$}][c]{\phantom{-}0} \\ 
  \mathmakebox[\widthof{$-1$}][c]{-1} & \mathmakebox[\widthof{$-1$}][c]{\phantom{-}0} & \mathmakebox[\widthof{$-1$}][c]{\phantom{-}0} & \mathmakebox[\widthof{$-1$}][c]{\phantom{-}1} \\ 
  \mathmakebox[\widthof{$-1$}][c]{\phantom{-}0} & \mathmakebox[\widthof{$-1$}][c]{-1} & \mathmakebox[\widthof{$-1$}][c]{\phantom{-}1} & \mathmakebox[\widthof{$-1$}][c]{\phantom{-}0} \\ 
  \mathmakebox[\widthof{$-1$}][c]{\phantom{-}0} & \mathmakebox[\widthof{$-1$}][c]{-1} & \mathmakebox[\widthof{$-1$}][c]{\phantom{-}0} & \mathmakebox[\widthof{$-1$}][c]{\phantom{-}1} \\ 
  \mathmakebox[\widthof{$-1$}][c]{\phantom{-}0} & \mathmakebox[\widthof{$-1$}][c]{\phantom{-}0} & \mathmakebox[\widthof{$-1$}][c]{-1} & \mathmakebox[\widthof{$-1$}][c]{\phantom{-}1}

    \end{pmatrix}.
\end{align}
Substituting these matrices into Equation \eqref{eq:H_bot} gives the aggregate hat matrix,
\begin{align}\label{eq:H_bot_eg}
    \bm{H}_\bot = \begin{pmatrix}
        % latex table generated in R 4.4.1 by xtable 1.8-4 package
% Mon May 19 17:30:36 2025
 \mathmakebox[\widthof{$-1.000$}][c]{\phantom{-}0.812} & \mathmakebox[\widthof{$-1.000$}][c]{\phantom{-}0.113} & \mathmakebox[\widthof{$-1.000$}][c]{\phantom{-}0.074} & \mathmakebox[\widthof{$-1.000$}][c]{-0.134} & \mathmakebox[\widthof{$-1.000$}][c]{-0.054} & \mathmakebox[\widthof{$-1.000$}][c]{-0.020} \\ 
  \mathmakebox[\widthof{$-1.000$}][c]{\phantom{-}0.375} & \mathmakebox[\widthof{$-1.000$}][c]{\phantom{-}0.424} & \mathmakebox[\widthof{$-1.000$}][c]{\phantom{-}0.201} & \mathmakebox[\widthof{$-1.000$}][c]{\phantom{-}0.313} & \mathmakebox[\widthof{$-1.000$}][c]{\phantom{-}0.061} & \mathmakebox[\widthof{$-1.000$}][c]{-0.262} \\ 
  \mathmakebox[\widthof{$-1.000$}][c]{\phantom{-}0.337} & \mathmakebox[\widthof{$-1.000$}][c]{\phantom{-}0.277} & \mathmakebox[\widthof{$-1.000$}][c]{\phantom{-}0.386} & \mathmakebox[\widthof{$-1.000$}][c]{\phantom{-}0.176} & \mathmakebox[\widthof{$-1.000$}][c]{\phantom{-}0.161} & \mathmakebox[\widthof{$-1.000$}][c]{\phantom{-}0.453}

    \end{pmatrix},
\end{align}
where each column corresponds to an edge $\in E_{\bot}$ and each row represents a comparison with the baseline treatment; $ab, ac$, and $ad$.

The first row of $\bm{H}_\bot$ in Equation (\ref{eq:H_bot_eg}) corresponds to the comparison $ab$ and defines an evidence flow network from node $a$ to node $b$, shown in panel (B) of Figure \ref{fig:EG-uni}. 
The thickness of each edge is proportional to the magnitude of the coefficient associated with that edge and the direction is given by its sign. 
For example, the final element of the row represents edge $[c,d]$.
The magnitude of flow through this edge is $0.02$ and the negative sign indicates that evidence flows in the direction from $d$ to $c$. 
In Figure \ref{fig:EG-uni}~(B) we observe the greatest flow along the edge from $a$ to $b$.
This indicates that the direct evidence on $ab$, $\hat\theta_{ab}^{\text{dir}}$, has the largest coefficient in the linear equation for $\hat\theta_{ab}$ and therefore contributes the most to the network estimate.

\subsubsection{Random walk on the unipartite graph}

In \cite{Davies:2022}, we showed that evidence flow has an interpretation in terms of a random walk. 
Consider a random walker moving on the unipartite NMA graph.
The probability of the walker hopping from node $v_j$ to node $v_k$ at any time step is proportional to the weight $w_{jk}$ of the direct estimate associated with that edge,
\begin{align}\label{eq:Tjk}
    T_{jk} = \frac{w_{jk}}{\sum_{l\neq j} w_{jl}}.
\end{align}
If there is no direct comparison between the two treatments represented by these nodes, then no hop can occur. 
We set $T_{jj}=0$ for all $j$ and write $\bm{T}$ for the transition matrix defined by these probabilities. 

For a particular comparison of interest $ab$, node $a$ is the initial state (source) of the walk and node $b$ is the absorbing state (sink). The walker starts its journey at node $a$ and hops around the network according to the transition probabilities in Equation (\ref{eq:Tjk}) until it reaches node $b$. The random walk interpretation of the flow of evidence is then as follows: 
\begin{theorem}\label{th:rw-flow}
For the comparison of treatments $a$ and $b$, the element of the aggregate hat matrix $\bm{H}_\bot$ that defines the flow of evidence through the direct comparison $jk$ is equal to the expected net number of times\footnotemark[3] a random walker starting at node $a$ on the unipartite NMA graph moves along the edge from $v_j$ to $v_k$ before it reaches node $b$.  
\end{theorem}
\footnotetext[3]{The net number of times the walker moves from $v_j$ to $v_k$ is the number of crossings in the direction from $v_j$ to $v_k$ minus the number of crossings in the opposite direction.} 
In \cite{Davies:2022} we proved this theorem using the pre-established analogies between (i) random walks and electrical networks (\cite{DoyleSnell:1984}) and (ii) electrical networks and NMA (\cite{Rucker:2012}). 
We refer to the original article for details.

\subsection{Bipartite evidence flow}
The evidence flow network on the unipartite graph visualises how the direct evidence combines to give the network estimates.
\cite{Papakon:2018} showed that these flows can be used to calculate the proportion contributions of different paths of evidence.
However, it is not clear how this relates to the contributions of individual trials. 
In this section, we instead use the arm-level hat matrix to construct an evidence flow network on the \textit{bipartite} NMA graph. 

\subsubsection{The arm level hat matrix}\label{sec:arm-hat-flow}
The arm-level hat matrix $\bm{H}_{\top\bot}$ contains the coefficients of a linear equation relating the treatment effect estimates to the observations in each arm of each trial.
Each column $ij$ of $\bm{H}_{\top\bot}$ corresponds to an edge connecting trial node $u_i$ to treatment node $v_j$ in the bipartite graph. 
Each row $kl$ corresponds to a comparison between two treatments, represented by bottom nodes $v_k$ and $v_l$. 

Similar to the aggregate hat matrix, the elements of each row of $\bm{H}_{\top\bot}$ have the properties of a flow network.
We write $H_{\top\bot, ij}^{(kl)}$ for the element in row $kl$ and column $ij$. 
We keep to our convention that $i$ labels trials (top nodes) while indices $j,k,l$ label treatments (bottom nodes). 
For a particular comparison $kl$, the arm-level hat matrix elements have the following properties: 
\begin{enumerate}
    \item the sum of outflows from node $v_k$ is equal to one, $\sum_{i} H_{\top\bot, ki}^{(kl)}=1$,
    \item the sum of inflows to node $v_l$ is equal to one, $\sum_{i} H_{\top\bot, il}^{(kl)}=1$,
    \item and at every intermediate node, the sum of outflows equals the sum of inflows. This is true for treatment nodes $v_j \neq v_k,v_l$, $$\sum_{i} H_{\top\bot, ji}^{(kl)}=\sum_{i} H_{\top\bot, ij}^{(kl)}$$ and all trial nodes $u_i\in V_\top$,  $$\sum_{j} H_{\top\bot, ij}^{(kl)}=\sum_{j} H_{\top\bot, ji}^{(kl)}$$
\end{enumerate}
Therefore, each row $kl$ of $\bm{H}_{\top\bot}$ describes a directed network of flows on the bipartite graph that starts at node $v_k$ and ends at node $v_l$. 
Due to the bipartite structure, edges are directed either from treatment nodes to trial nodes or vice versa. There can be no flow between two treatment nodes or between two trial nodes. 
The magnitude of flow between trial node $u_i$ and treatment node $v_j$ is given by the magnitude of the arm-level hat matrix coefficient $H_{\top\bot, ij}^{(kl)}$ associated with the observation in arm $t_{i,\ell}=j$ of trial $i$. 
The direction of flow is indicated by its sign.

\begin{figure}[b]
    \centering
    \includegraphics[width=0.8\linewidth]{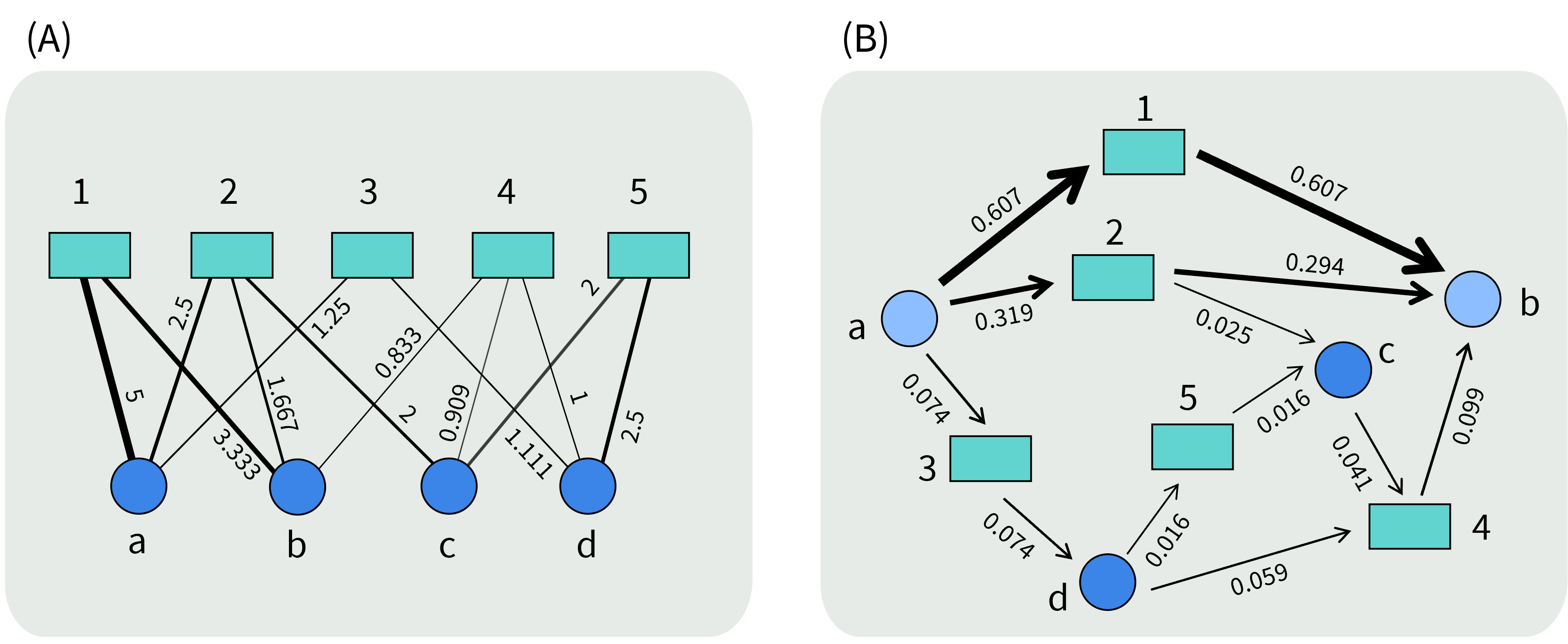}
    \caption{(A) Weighted bipartite graph for the fictional example network described in Table \ref{Tab:EG}. Rectangular nodes represent trials and circular nodes represent treatments. Each edge represents an arm of a trial. The thickness of an edge $[u_i,v_j]$ is proportional to the CE inverse-variance weight associated with that arm, $\sigma_{i,j}^{-2}$. These weights are shown in the edge labels. (B) The bipartite evidence flow network for the comparison of treatments $a$ and $b$. The thickness of each edge is proportional to the corresponding element of the arm-level hat matrix $\bm{H}_{\top\bot}$ shown in the edge labels. The direction of each edge is given by the sign of that element. }
    \label{fig:EG-bi}
\end{figure}

The weight associated with each arm-level observation $\mu_{i,j}$ is given by the inverse of its variance. 
For the CE model this is $\sigma_{i,j}^{-2}$ and for the RE model, $(\sigma_{i,j}^2+\frac{\tau^2}{2})^{-1}$. 
These are the weights associated with each edge in the bipartite graph.
Panel (A) of Figure \ref{fig:EG-bi} shows the weighted bipartite graph for the fictional example in Table \ref{Tab:EG}. 
The graph has $K_{\top\bot}=12$ edges connecting each trial to the treatments it compares, $E_{\top\bot} = [1a,\linebreak[1] 1b,\linebreak[1] 2a,\linebreak[1] 2b,\linebreak[1] 2c,\linebreak[1] 3a,\linebreak[1] 3d,\linebreak[1] 4b,\linebreak[1] 4c,\linebreak[1] 4d,\linebreak[1] 5c,\linebreak[1] 5d]$. 
The thickness of each edge is proportional to the CE weight of that trial arm calculated from the inverse arm-level variances in Table \ref{Tab:EG}. 
These weights are shown in the edge labels. 
The corresponding weight matrix $\bm{W}=\bm{\Sigma}^{-1}$ is defined on the contrast-level and is given by
\begin{align}
\bm{W}=\begin{pmatrix}
% latex table generated in R 4.4.1 by xtable 1.8-4 package
% Mon May 19 17:07:36 2025
 \mathmakebox[\widthof{$-1.000$}][c]{\phantom{-}2.000} & \mathmakebox[\widthof{$-1.000$}][c]{\phantom{-}0} & \mathmakebox[\widthof{$-1.000$}][c]{\phantom{-}0} & \mathmakebox[\widthof{$-1.000$}][c]{\phantom{-}0} & \mathmakebox[\widthof{$-1.000$}][c]{\phantom{-}0} & \mathmakebox[\widthof{$-1.000$}][c]{\phantom{-}0} & \mathmakebox[\widthof{$-1.000$}][c]{\phantom{-}0} \\ 
  \mathmakebox[\widthof{$-1.000$}][c]{\phantom{-}0} & \mathmakebox[\widthof{$-1.000$}][c]{\phantom{-}1.216} & \mathmakebox[\widthof{$-1.000$}][c]{-0.541} & \mathmakebox[\widthof{$-1.000$}][c]{\phantom{-}0} & \mathmakebox[\widthof{$-1.000$}][c]{\phantom{-}0} & \mathmakebox[\widthof{$-1.000$}][c]{\phantom{-}0} & \mathmakebox[\widthof{$-1.000$}][c]{\phantom{-}0} \\ 
  \mathmakebox[\widthof{$-1.000$}][c]{\phantom{-}0} & \mathmakebox[\widthof{$-1.000$}][c]{-0.541} & \mathmakebox[\widthof{$-1.000$}][c]{\phantom{-}1.351} & \mathmakebox[\widthof{$-1.000$}][c]{\phantom{-}0} & \mathmakebox[\widthof{$-1.000$}][c]{\phantom{-}0} & \mathmakebox[\widthof{$-1.000$}][c]{\phantom{-}0} & \mathmakebox[\widthof{$-1.000$}][c]{\phantom{-}0} \\ 
  \mathmakebox[\widthof{$-1.000$}][c]{\phantom{-}0} & \mathmakebox[\widthof{$-1.000$}][c]{\phantom{-}0} & \mathmakebox[\widthof{$-1.000$}][c]{\phantom{-}0} & \mathmakebox[\widthof{$-1.000$}][c]{\phantom{-}0.588} & \mathmakebox[\widthof{$-1.000$}][c]{\phantom{-}0} & \mathmakebox[\widthof{$-1.000$}][c]{\phantom{-}0} & \mathmakebox[\widthof{$-1.000$}][c]{\phantom{-}0} \\ 
  \mathmakebox[\widthof{$-1.000$}][c]{\phantom{-}0} & \mathmakebox[\widthof{$-1.000$}][c]{\phantom{-}0} & \mathmakebox[\widthof{$-1.000$}][c]{\phantom{-}0} & \mathmakebox[\widthof{$-1.000$}][c]{\phantom{-}0} & \mathmakebox[\widthof{$-1.000$}][c]{\phantom{-}0.608} & \mathmakebox[\widthof{$-1.000$}][c]{-0.331} & \mathmakebox[\widthof{$-1.000$}][c]{\phantom{-}0} \\ 
  \mathmakebox[\widthof{$-1.000$}][c]{\phantom{-}0} & \mathmakebox[\widthof{$-1.000$}][c]{\phantom{-}0} & \mathmakebox[\widthof{$-1.000$}][c]{\phantom{-}0} & \mathmakebox[\widthof{$-1.000$}][c]{\phantom{-}0} & \mathmakebox[\widthof{$-1.000$}][c]{-0.331} & \mathmakebox[\widthof{$-1.000$}][c]{\phantom{-}0.635} & \mathmakebox[\widthof{$-1.000$}][c]{\phantom{-}0} \\ 
  \mathmakebox[\widthof{$-1.000$}][c]{\phantom{-}0} & \mathmakebox[\widthof{$-1.000$}][c]{\phantom{-}0} & \mathmakebox[\widthof{$-1.000$}][c]{\phantom{-}0} & \mathmakebox[\widthof{$-1.000$}][c]{\phantom{-}0} & \mathmakebox[\widthof{$-1.000$}][c]{\phantom{-}0} & \mathmakebox[\widthof{$-1.000$}][c]{\phantom{-}0} & \mathmakebox[\widthof{$-1.000$}][c]{\phantom{-}1.111}

\end{pmatrix},
\end{align}
where each row and column represents a comparison to the trial-specific baseline in each trial. The design matrix of the network is
\begin{align}
    \bm{X}= \begin{pmatrix}
        % latex table generated in R 4.4.1 by xtable 1.8-4 package
% Mon May 19 17:07:31 2025
 \mathmakebox[\widthof{$-1$}][c]{\phantom{-}1} & \mathmakebox[\widthof{$-1$}][c]{\phantom{-}0} & \mathmakebox[\widthof{$-1$}][c]{\phantom{-}0} \\ 
  \mathmakebox[\widthof{$-1$}][c]{\phantom{-}1} & \mathmakebox[\widthof{$-1$}][c]{\phantom{-}0} & \mathmakebox[\widthof{$-1$}][c]{\phantom{-}0} \\ 
  \mathmakebox[\widthof{$-1$}][c]{\phantom{-}0} & \mathmakebox[\widthof{$-1$}][c]{\phantom{-}1} & \mathmakebox[\widthof{$-1$}][c]{\phantom{-}0} \\ 
  \mathmakebox[\widthof{$-1$}][c]{\phantom{-}0} & \mathmakebox[\widthof{$-1$}][c]{\phantom{-}0} & \mathmakebox[\widthof{$-1$}][c]{\phantom{-}1} \\ 
  \mathmakebox[\widthof{$-1$}][c]{-1} & \mathmakebox[\widthof{$-1$}][c]{\phantom{-}1} & \mathmakebox[\widthof{$-1$}][c]{\phantom{-}0} \\ 
  \mathmakebox[\widthof{$-1$}][c]{-1} & \mathmakebox[\widthof{$-1$}][c]{\phantom{-}0} & \mathmakebox[\widthof{$-1$}][c]{\phantom{-}1} \\ 
  \mathmakebox[\widthof{$-1$}][c]{\phantom{-}0} & \mathmakebox[\widthof{$-1$}][c]{-1} & \mathmakebox[\widthof{$-1$}][c]{\phantom{-}1}

    \end{pmatrix},
\end{align}
and the matrix relating the arm-level and contrast-level observations is
\begin{align}
    \bm{C} = \begin{pmatrix}
        % latex table generated in R 4.4.1 by xtable 1.8-4 package
% Mon May 19 17:07:34 2025
 \mathmakebox[\widthof{$-1$}][c]{-1} & \mathmakebox[\widthof{$-1$}][c]{\phantom{-}1} & \mathmakebox[\widthof{$-1$}][c]{\phantom{-}0} & \mathmakebox[\widthof{$-1$}][c]{\phantom{-}0} & \mathmakebox[\widthof{$-1$}][c]{\phantom{-}0} & \mathmakebox[\widthof{$-1$}][c]{\phantom{-}0} & \mathmakebox[\widthof{$-1$}][c]{\phantom{-}0} & \mathmakebox[\widthof{$-1$}][c]{\phantom{-}0} & \mathmakebox[\widthof{$-1$}][c]{\phantom{-}0} & \mathmakebox[\widthof{$-1$}][c]{\phantom{-}0} & \mathmakebox[\widthof{$-1$}][c]{\phantom{-}0} & \mathmakebox[\widthof{$-1$}][c]{\phantom{-}0} \\ 
  \mathmakebox[\widthof{$-1$}][c]{\phantom{-}0} & \mathmakebox[\widthof{$-1$}][c]{\phantom{-}0} & \mathmakebox[\widthof{$-1$}][c]{-1} & \mathmakebox[\widthof{$-1$}][c]{\phantom{-}1} & \mathmakebox[\widthof{$-1$}][c]{\phantom{-}0} & \mathmakebox[\widthof{$-1$}][c]{\phantom{-}0} & \mathmakebox[\widthof{$-1$}][c]{\phantom{-}0} & \mathmakebox[\widthof{$-1$}][c]{\phantom{-}0} & \mathmakebox[\widthof{$-1$}][c]{\phantom{-}0} & \mathmakebox[\widthof{$-1$}][c]{\phantom{-}0} & \mathmakebox[\widthof{$-1$}][c]{\phantom{-}0} & \mathmakebox[\widthof{$-1$}][c]{\phantom{-}0} \\ 
  \mathmakebox[\widthof{$-1$}][c]{\phantom{-}0} & \mathmakebox[\widthof{$-1$}][c]{\phantom{-}0} & \mathmakebox[\widthof{$-1$}][c]{-1} & \mathmakebox[\widthof{$-1$}][c]{\phantom{-}0} & \mathmakebox[\widthof{$-1$}][c]{\phantom{-}1} & \mathmakebox[\widthof{$-1$}][c]{\phantom{-}0} & \mathmakebox[\widthof{$-1$}][c]{\phantom{-}0} & \mathmakebox[\widthof{$-1$}][c]{\phantom{-}0} & \mathmakebox[\widthof{$-1$}][c]{\phantom{-}0} & \mathmakebox[\widthof{$-1$}][c]{\phantom{-}0} & \mathmakebox[\widthof{$-1$}][c]{\phantom{-}0} & \mathmakebox[\widthof{$-1$}][c]{\phantom{-}0} \\ 
  \mathmakebox[\widthof{$-1$}][c]{\phantom{-}0} & \mathmakebox[\widthof{$-1$}][c]{\phantom{-}0} & \mathmakebox[\widthof{$-1$}][c]{\phantom{-}0} & \mathmakebox[\widthof{$-1$}][c]{\phantom{-}0} & \mathmakebox[\widthof{$-1$}][c]{\phantom{-}0} & \mathmakebox[\widthof{$-1$}][c]{-1} & \mathmakebox[\widthof{$-1$}][c]{\phantom{-}1} & \mathmakebox[\widthof{$-1$}][c]{\phantom{-}0} & \mathmakebox[\widthof{$-1$}][c]{\phantom{-}0} & \mathmakebox[\widthof{$-1$}][c]{\phantom{-}0} & \mathmakebox[\widthof{$-1$}][c]{\phantom{-}0} & \mathmakebox[\widthof{$-1$}][c]{\phantom{-}0} \\ 
  \mathmakebox[\widthof{$-1$}][c]{\phantom{-}0} & \mathmakebox[\widthof{$-1$}][c]{\phantom{-}0} & \mathmakebox[\widthof{$-1$}][c]{\phantom{-}0} & \mathmakebox[\widthof{$-1$}][c]{\phantom{-}0} & \mathmakebox[\widthof{$-1$}][c]{\phantom{-}0} & \mathmakebox[\widthof{$-1$}][c]{\phantom{-}0} & \mathmakebox[\widthof{$-1$}][c]{\phantom{-}0} & \mathmakebox[\widthof{$-1$}][c]{-1} & \mathmakebox[\widthof{$-1$}][c]{\phantom{-}1} & \mathmakebox[\widthof{$-1$}][c]{\phantom{-}0} & \mathmakebox[\widthof{$-1$}][c]{\phantom{-}0} & \mathmakebox[\widthof{$-1$}][c]{\phantom{-}0} \\ 
  \mathmakebox[\widthof{$-1$}][c]{\phantom{-}0} & \mathmakebox[\widthof{$-1$}][c]{\phantom{-}0} & \mathmakebox[\widthof{$-1$}][c]{\phantom{-}0} & \mathmakebox[\widthof{$-1$}][c]{\phantom{-}0} & \mathmakebox[\widthof{$-1$}][c]{\phantom{-}0} & \mathmakebox[\widthof{$-1$}][c]{\phantom{-}0} & \mathmakebox[\widthof{$-1$}][c]{\phantom{-}0} & \mathmakebox[\widthof{$-1$}][c]{-1} & \mathmakebox[\widthof{$-1$}][c]{\phantom{-}0} & \mathmakebox[\widthof{$-1$}][c]{\phantom{-}1} & \mathmakebox[\widthof{$-1$}][c]{\phantom{-}0} & \mathmakebox[\widthof{$-1$}][c]{\phantom{-}0} \\ 
  \mathmakebox[\widthof{$-1$}][c]{\phantom{-}0} & \mathmakebox[\widthof{$-1$}][c]{\phantom{-}0} & \mathmakebox[\widthof{$-1$}][c]{\phantom{-}0} & \mathmakebox[\widthof{$-1$}][c]{\phantom{-}0} & \mathmakebox[\widthof{$-1$}][c]{\phantom{-}0} & \mathmakebox[\widthof{$-1$}][c]{\phantom{-}0} & \mathmakebox[\widthof{$-1$}][c]{\phantom{-}0} & \mathmakebox[\widthof{$-1$}][c]{\phantom{-}0} & \mathmakebox[\widthof{$-1$}][c]{\phantom{-}0} & \mathmakebox[\widthof{$-1$}][c]{\phantom{-}0} & \mathmakebox[\widthof{$-1$}][c]{-1} & \mathmakebox[\widthof{$-1$}][c]{\phantom{-}1}

    \end{pmatrix}.
\end{align}
Substituting these matrices into Equation (\ref{eq:H-bi}) yields the arm-level hat matrix,
\begin{equation*}
\bm{H}_{\top\bot} = \begin{pmatrix}
% latex table generated in R 4.4.1 by xtable 1.8-4 package
% Mon May 19 17:07:38 2025
 \mathmakebox[\widthof{$-1.000$}][c]{-0.607} & \mathmakebox[\widthof{$-1.000$}][c]{\phantom{-}0.607} & \mathmakebox[\widthof{$-1.000$}][c]{-0.319} & \mathmakebox[\widthof{$-1.000$}][c]{\phantom{-}0.294} & \mathmakebox[\widthof{$-1.000$}][c]{\phantom{-}0.025} & \mathmakebox[\widthof{$-1.000$}][c]{-0.074} & \mathmakebox[\widthof{$-1.000$}][c]{\phantom{-}0.074} & \mathmakebox[\widthof{$-1.000$}][c]{\phantom{-}0.099} & \mathmakebox[\widthof{$-1.000$}][c]{-0.041} & \mathmakebox[\widthof{$-1.000$}][c]{-0.059} & \mathmakebox[\widthof{$-1.000$}][c]{\phantom{-}0.016} & \mathmakebox[\widthof{$-1.000$}][c]{-0.016} \\ 
  \mathmakebox[\widthof{$-1.000$}][c]{-0.280} & \mathmakebox[\widthof{$-1.000$}][c]{\phantom{-}0.280} & \mathmakebox[\widthof{$-1.000$}][c]{-0.519} & \mathmakebox[\widthof{$-1.000$}][c]{-0.113} & \mathmakebox[\widthof{$-1.000$}][c]{\phantom{-}0.632} & \mathmakebox[\widthof{$-1.000$}][c]{-0.201} & \mathmakebox[\widthof{$-1.000$}][c]{\phantom{-}0.201} & \mathmakebox[\widthof{$-1.000$}][c]{-0.167} & \mathmakebox[\widthof{$-1.000$}][c]{\phantom{-}0.166} & \mathmakebox[\widthof{$-1.000$}][c]{\phantom{-}0.001} & \mathmakebox[\widthof{$-1.000$}][c]{\phantom{-}0.202} & \mathmakebox[\widthof{$-1.000$}][c]{-0.202} \\ 
  \mathmakebox[\widthof{$-1.000$}][c]{-0.252} & \mathmakebox[\widthof{$-1.000$}][c]{\phantom{-}0.252} & \mathmakebox[\widthof{$-1.000$}][c]{-0.362} & \mathmakebox[\widthof{$-1.000$}][c]{-0.031} & \mathmakebox[\widthof{$-1.000$}][c]{\phantom{-}0.394} & \mathmakebox[\widthof{$-1.000$}][c]{-0.386} & \mathmakebox[\widthof{$-1.000$}][c]{\phantom{-}0.386} & \mathmakebox[\widthof{$-1.000$}][c]{-0.221} & \mathmakebox[\widthof{$-1.000$}][c]{-0.045} & \mathmakebox[\widthof{$-1.000$}][c]{\phantom{-}0.265} & \mathmakebox[\widthof{$-1.000$}][c]{-0.349} & \mathmakebox[\widthof{$-1.000$}][c]{\phantom{-}0.349}

\end{pmatrix},
\end{equation*}
where each column corresponds to an edge $\in E_{\top\bot}$ and each row represents a comparison to the baseline; $ab, ac$, and  $ad$.

The first row of $\bm{H}_{\top\bot}$ corresponds to comparison $ab$ and describes a flow network on the bipartite graph from bottom node $a$ to bottom node $b$.
This is shown in panel (B) of Figure \ref{fig:EG-bi}.
The thickness of each edge is proportional to the magnitude of the associated coefficient and the direction is given by its sign. Unlike the unipartite flow network in Figure \ref{fig:EG-uni}~(B), the bipartite representation shows how the evidence flows through the individual trials. For example, we see that the direct evidence comes from trials 1 and 2 with trial 1 exerting the most influence. The flow through the other three trial nodes is weaker, indicating that they make a smaller contribution to this comparison. 

\subsubsection{Random walk on the bipartite graph}\label{sec:RW-bi}

Inspired by \cite{Zeng:2024}'s recent work on higher order random walk (HoRW) models, we construct a random walk on the bipartite NMA graph. 
To do so, we define a weighted version of the biadjacency matrix $\boldsymbol{B}$. 
Now, each element $B_{ij}$ indicates not only the presence of edge $[u_i,v_j]$, but also its (inverse-variance) weight. 
For the RE model, we have
\begin{align}\label{eq:Bweight}
    B_{ij} = \begin{cases}
        (\sigma_{i,j}^2+\frac{\tau^2}{2})^{-1} & \text{ if } j\in\mathcal{T}_i\\
        0 & \text{otherwise},
    \end{cases}
\end{align}
such that $B_{ij}> 0$ indicates that treatment $j$ appears in trial $i$. For the CE model we set $\tau=0$. 

\begin{figure}
    \centering
    \includegraphics[width=0.4\linewidth]{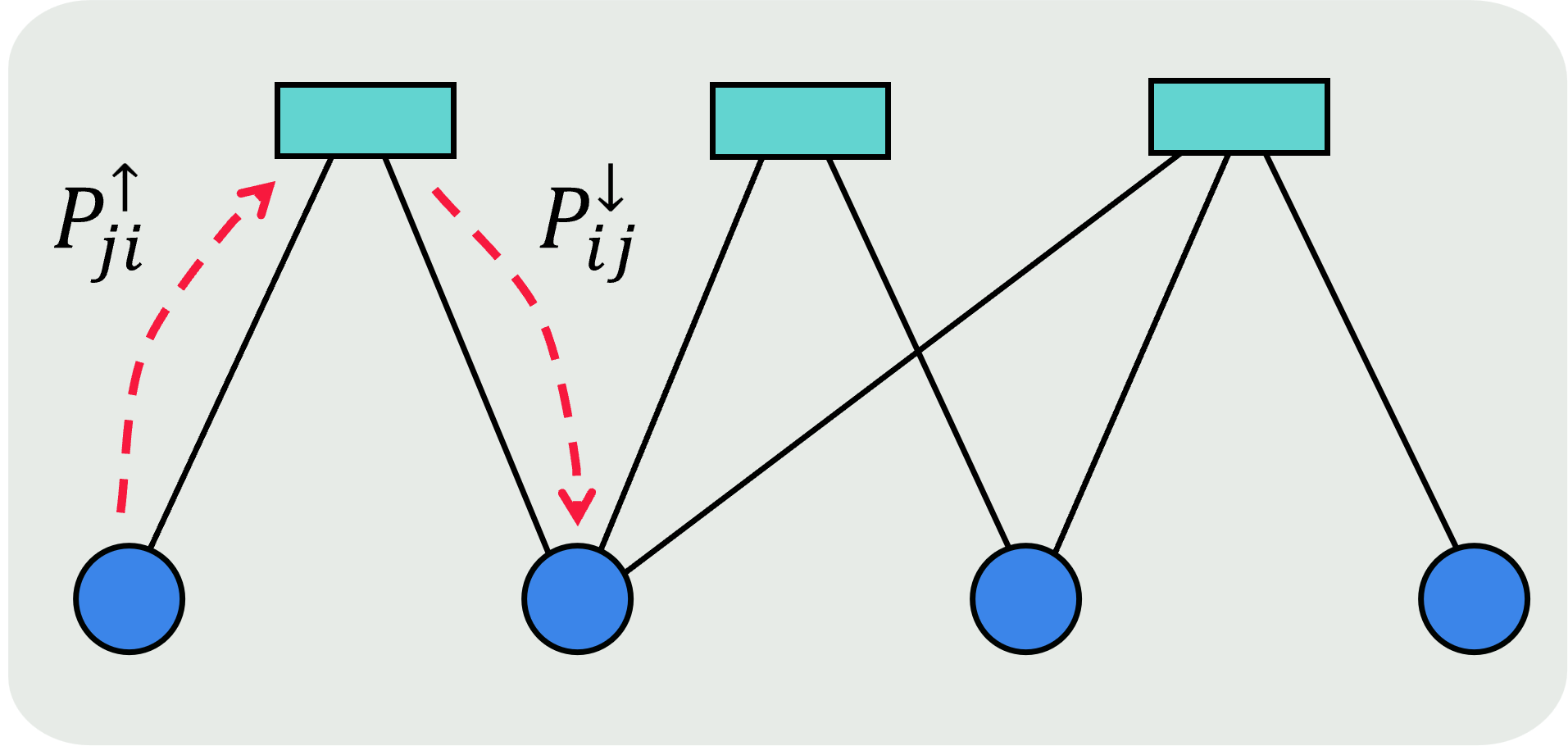}
    \caption{An illustration of the higher order random walk (HoRW) on a bipartite graph. The walk consists of two parts: an upward walk from a bottom node $v_j$ to a top node $u_i$ with probability $P_{ji}^{\uparrow}$, and a downward walk from a top node $u_i$ to a bottom node $v_j$ with probability $P_{ij}^{\downarrow}$.}
    \label{fig:UpDownWalk}
\end{figure}

On the bipartite graph, a walker can only hop between top and bottom nodes. As shown in Figure \ref{fig:UpDownWalk}, the HoRW is therefore made up of two parts: a downward walk from top to bottom, and an upward walk from bottom to top. We define the transition probabilities of these walks in terms of $\bm{B}$.

The downward walk involves a hop from a trial (top) node to a treatment (bottom) node. The transition probability of moving along an edge connecting two nodes is proportional to the weight associated with the corresponding trial arm. More specifically, the probability of hopping from trial node $u_i$ to treatment node $v_j$ at any time step is 
\begin{align}\label{eq:Pdown}
    P^{\downarrow}_{ij} = \frac{B_{ij}}{\sum_{j'=1}^{N} B_{ij'}},
\end{align}
where the denominator is equal to the weighted degree of $u_i$, $\sum_{j'=1}^{N} B_{ij'} = \sum_{\ell=1}^{n_i} (\sigma_{i,t_{i\ell}}^2+\frac{\tau^2}{2})^{-1}$. The $M\times N$ downstream transition matrix is then
\begin{align}\label{eq:Pdown-mat}
    \bm{P}^{\downarrow} = \bm{\Lambda}_{\top}^{-1}\bm{B},
\end{align}
where $\bm{\Lambda}_{\top}$ is an $M\times M$ diagonal matrix whose diagonal elements are the weighted degrees of the trial nodes.

The upward walk is characterised by a hop from a treatment (bottom) node to a trial (top) node. As before, transition probabilities are proportional to the weights of the trial arms. This time, however, normalization is performed with respect to the weighted degree of the \textit{treatment} node. The probability of hopping from treatment node $v_j$ to trial node $u_i$ is then
\begin{align}\label{eq:Pup}
    P^{\uparrow}_{ji} = \frac{B_{ij}}{\sum_{i'=1}^{M} B_{i'j}},
\end{align}
where $\sum_{i'=1}^{M} B_{i'j}=\sum_{i' \in \mathcal{M}_j} (\sigma_{i',j}^2+\frac{\tau^2}{2})^{-1}$ and $\mathcal{M}_j=[i'|\mathcal{T}_{i'} \supseteq j]$ is the set of trials containing treatment $j$. 
This yields the upstream transition matrix,
\begin{align}\label{eq:Pup-mat}
    \bm{P}^{\uparrow} = \bm{\Lambda}_{\bot}^{-1}\bm{B}',
\end{align}
where $\bm{\Lambda}_{\bot}$ is an $N \times N$ diagonal matrix containing the weighted degrees of the treatment nodes.

The HoRW is fully described by the transition matrices $\bm{P}^{\uparrow}$ and $\bm{P}^{\downarrow}$. 
As before, for a given treatment comparison $ab$, we define (bottom) node $a$ as the initial state and (bottom) node $b$ as the absorbing state. 
The walker starts at $a$ and hops between treatment and trial nodes according to $\bm{P}^{\uparrow}$ and $\bm{P}^{\downarrow}$ until it reaches $b$. 
Due to the bipartite structure, the walker cannot hop between two treatment nodes or between two trial nodes.
Based on Theorem \ref{th:rw-flow}, we make the following conjecture relating this random walk to the bipartite evidence flow network: 
\begin{conj}\label{conj:rw-bi}
For the comparison of treatments $a$ and $b$, the element of the arm-level hat matrix $\bm{H}_{\top\bot}$ that defines the flow of evidence through arm $t_{i\ell}=j$ of trial $i$ is equal to the expected net number of times a random walker starting at node $a$ on the bipartite NMA graph moves along the edge from treatment node $v_j$ to trial node $u_i$ before it reaches node $b$. 
\end{conj}
To prove this conjecture, we turn again to the analogies between NMA, electrical networks, and random walks described in \cite{Davies:2022}. 
In Section \ref{app:electric} of the Supplementary Material, we extend these analogies to the bipartite framework and the HoRW. 
We summarize the main ideas below.

The weight of an edge in an electrical network is equal to the inverse of the resistance in that edge.  
\cite{Rucker:2012} showed that these resistances are analogous to variances in NMA. 
Therefore, we interpret the variance associated with each trial arm as a resistance in the corresponding edge of the bipartite graph. 
The resulting matrix of edge resistances, $\bm{R}$, has dimensions $K_{\top\bot}\times K_{\top\bot}$ and contains the arm-level variances, $\sigma_{i,j}^2+\frac{\tau^2}{2}$, on its diagonal. It is related to the inverse-variance weight matrix via $\bm{W}^{-1} = \bm{C}\bm{R}\bm{C'}$. 
The analogy between electrical networks and random walks (\cite{DoyleSnell:1984}) equates the movement of the random walker with the electrical current flowing through each edge in the network. 
In electrical network theory, the currents in each edge are calculated via
\begin{align}\label{eq:Iprime}
    \bm{I'} = \bm{J'}(\bm{B}_{\top\bot}'\bm{R}^{-1}\bm{B}_{\top\bot})^{+}\bm{B}_{\top\bot}'\bm{R}^{-1},
\end{align}
where $\bm{J'}=[\bm{0}_{(N-1)\times M} \hspace{3pt} \bm{C}_N]$ is a matrix containing the currents at each node, $\bm{0}_{(N-1)\times M}$ is an $(N-1)\times M$ matrix of zeroes, and $\bm{B}_{\top\bot}$ is the oriented edge-vertex incidence matrix of the bipartite graph described in Section \ref{sec:bipartite}. 
To prove Conjecture \ref{conj:rw-bi}, it suffices to show that $\bm{I'}$ is equal to the arm-level hat matrix $\bm{H}_{\top\bot}$. We evaluate this conjecture empirically in Sections \ref{sec:application} and \ref{sec:simulations}.

\subsection{Unipartite flow from the bipartite random walk}
As a final step, we use the random walk on the \textit{bipartite} graph to obtain the original evidence flows on the \textit{unipartite} graph.
Following \cite{Zeng:2024}, we define a two-step random walk with an $N\times N$ transition matrix calculated from the upstream and downstream transition matrices,  
\begin{align}\label{eq:P}
    \bm{P} = \bm{P}^{\uparrow}\bm{P}^{\downarrow}.
\end{align}
Each element $P_{jk}$ describes the probability of two consecutive hops on the bipartite graph: a walker currently at bottom node $v_j$ hops to any top node in the first step, followed immediately by a hop to bottom node $v_k$ in the second step. 
The combination of these two consecutive steps defines a single transition (from  node $v_j$ to node $v_k$) on the unipartite projection of the bipartite graph.
Therefore, for any treatment node $v_j$ connected to a trial node $u_i$, there is always a non-zero probability for the walker to hop from $v_j$ to $u_i$ and then back to $v_j$ in a single transition. 
This means $P_{jj}>0$ for all $j$. 
In other words, the unipartite walk defined by $\bm{P}$ allows the walker to remain at the same node at a given time step. 

Given the random walk interpretation of evidence flow, we propose that the unipartite evidence flows described by the aggregate hat matrix $\bm{H}_\bot$ can be obtained from the bipartite random walk as follows:

\begin{conj}\label{conj:uni-bi}
    For the comparison of treatments $a$ and $b$, the element of the aggregate hat matrix $\bm{H_\bot}$ that defines the flow of evidence through the direct comparison $jk$ on the unipartite graph is equal to the expected net number of times a random walker starting at node $a$ on the bipartite graph moves from (bottom node) $v_j$ to (bottom node) $v_k$ in two consecutive steps (via any trial node) before it reaches node $b$. 
\end{conj}

In this conjecture, we are interested in the net number of times a walker moves along each edge as it travels from $a$ to $b$. 
Any transitions that involve staying at a node do not affect these counts. 
We can adapt the two-step transition matrix $\bm{P}$ to prohibit transitions to the same node; first setting the diagonal elements equal to zero, then renormalising the remaining probabilities in each row. We call the resulting matrix $\bm{\Tilde{P}}$. 
This matrix describes a random walk on the unipartite graph that does not allow transitions to the same node ($\Tilde{P}_{jj}=0$) but yields the same expected number of edge crossings as the two-step walk described by $\bm{P}$. 
In Theorem \ref{th:rw-flow}, we expressed the evidence flow in $\bm{H}_\bot$ in terms of a random walker moving on the unipartite graph according to the transition matrix $\bm{T}$ in Equation (\ref{eq:Tjk}).
By specifying $T_{jj}=0$, this walk prohibits transitions to the same node.
Therefore, to prove Conjecture \ref{conj:uni-bi} it suffices to show that the renormalised two-step transition matrix is equal to the unipartite matrix, $\bm{\Tilde{P}}=\bm{T}$. 
We evaluate this conjecture empirically in the following sections.

\section{Application to motivating data set}\label{sec:application}

As an illustration, we return to the plaque psoriasis dataset described in Section \ref{sec:eg}. 
The treatments and trials define two sets of nodes, 
\begin{align}
    (v_1,\hdots,v_7) &= (\text{ETN, IXE\_Q2W, IXE\_Q4W, PBO, SEC\_150, SEC\_300, UST}),\\
    (u_1, \hdots, u_9)&=(\text{CLEAR, ERASURE, FEATURE, FIXTURE, IXORA-S, JUNCTURE, UNCOVER-1,}\nonumber\\ &\hspace{250pt}\text{UNCOVER-2, UNCOVER-3}). 
\end{align} 
We focus on the primary comparison of interest between $v_2:$ IXE\_Q2W and $v_6:$ SEC\_300, and explore the flow of evidence between these nodes on both the unipartite and bipartite graphs.
In fitting the models, we choose common effects ($\tau=0$) and use etanercept ($v_1:$ ETN) as the baseline treatment. 

%%%%% FIG
\begin{figure}[b]
    \centering
    \includegraphics[width=1\linewidth]{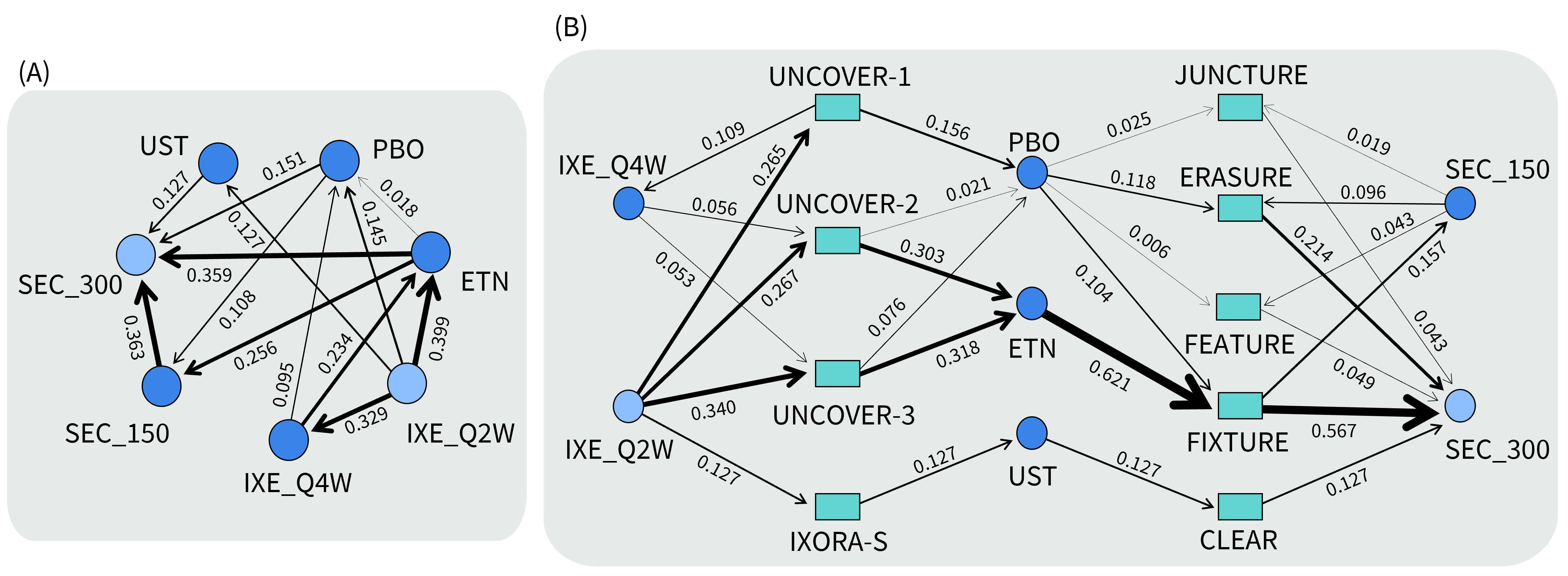}
    \caption{(A) A unipartite evidence flow network for the plaque psoriasis NMA. In this example, the comparison of interest is between secukinumab 300mg (SEC\_300) and ixekizumab every 2 weeks (IXE\_Q2W). Evidence flows from node IXE\_Q2W (the source) to node SEC\_300 (the sink) along edges representing direct comparisons between treatments. The thickness of each edge is proportional to the corresponding element of the aggregate hat matrix $\bm{H}_{\bot}$ shown in the edge labels. The direction of each edge is given by the sign of that element. (B) The bipartite evidence flow network for the same comparison. Again, node IXE\_Q2W is the source and node SEC\_300 is the sink but now edges represent the arms of the trials. The thickness and direction of each edge is given by the magnitude and sign of the corresponding element of the arm-level hat matrix $\bm{H}_{\top\bot}$.  }
    \label{fig:psoriasis-flow}
\end{figure}

\subsection{Unipartite evidence flow}
\subsubsection{The aggregate hat matrix}
We begin by applying the aggregate model to explore evidence flow on the unipartite graph.
The aggregate hat matrix $\boldsymbol{H}_\bot$ for the psoriasis dataset is shown in Appendix \ref{app:psoriasis}, along with its component matrices $\bm{B}_\bot$ and $\bm{W}_\bot$. 
Each row of $\bm{H}_\bot$ represents a comparison with the baseline, and each column represents an edge in the unipartite graph in Figure \ref{fig:psoriasis}~(A). 
Columns are ordered according to the standard convention $([v_1,v_2], [v_1,v_3], \hdots, [v_1,v_N], [v_2,v_3], [v_2,v_4],\allowbreak\hdots,[v_2,v_N],\hdots, [v_{N-1},v_N])$. 
If there is no direct evidence on the comparison of treatments $j$ and $k$, then there is no edge $[v_j,v_k]$ and that column is missing from $\bm{H}_\bot$.

Since the comparison of interest (IXE\_Q2W vs SEC\_300) does not involve the baseline treatment (ETN), we calculate the relevant hat matrix elements using the consistency relations in Equation (\ref{eq:hat-consistency}).
Subtracting the row corresponding to the comparison of IXE\_Q2W with the baseline from the row corresponding to the comparison of SEC\_300 with the baseline gives
\begin{align*}
     \bm{H}_{\bot}^{(\text{IXE\_Q2W, SEC\_300})}&=\bm{H}_{\bot}^{(\text{ETN, SEC\_300})}-\bm{H}_{\bot}^{(\text{ETN, IXE\_Q2W})} \nonumber\\
     &=(-0.399, -0.234, 0.018, 0.256, 0.359, 0.329, 0.145, 0.127, 0.095, 0.108, 0.151, 0.363, -0.127).
\end{align*}
As shown in panel (A) of Figure \ref{fig:psoriasis-flow}, these values define an evidence flow network on the unipartite graph with a source at node IXE\_Q2W and a sink at node SEC\_300. 
The absence of an edge between these two nodes indicates that no trials have directly compared these treatments. 
Instead, most of the evidence on this comparison comes indirectly via node ETN. 
Looking at the original unipartite graph in Figure \ref{fig:psoriasis}~(A), we might have expected that comparisons to placebo (PBO) would have a greater influence, since there are more trials involving this node. 
To explore this more thoroughly, we need to inspect the influence of individual trials and their arms.

\subsubsection{Random walk on the unipartite graph}
Using the aggregate weight matrix $\bm{W}_\bot$, we define a random walk on the unipartite graph in Figure \ref{fig:psoriasis}~(A). 
The transition matrix of this walk, calculated via Equation (\ref{eq:Tjk}), is 
\begin{align}\label{eq:T-psor}
\bm{T} &= \begin{pmatrix}
                    % latex table generated in R 4.4.1 by xtable 1.8-4 package
% Wed Feb 26 16:14:26 2025
 \mathmakebox[\widthof{-1.000}+\arraycolsep][c]{\phantom{-}0} & \mathmakebox[\widthof{-1.000}+\arraycolsep][c]{\phantom{-}0.208} & \mathmakebox[\widthof{-1.000}+\arraycolsep][c]{\phantom{-}0.363} & \mathmakebox[\widthof{-1.000}+\arraycolsep][c]{\phantom{-}0.110} & \mathmakebox[\widthof{-1.000}+\arraycolsep][c]{\phantom{-}0.178} & \mathmakebox[\widthof{-1.000}+\arraycolsep][c]{\phantom{-}0.141} & \mathmakebox[\widthof{-1.000}+\arraycolsep][c]{\phantom{-}0} \\ 
  \mathmakebox[\widthof{-1.000}+\arraycolsep][c]{\phantom{-}0.363} & \mathmakebox[\widthof{-1.000}+\arraycolsep][c]{\phantom{-}0} & \mathmakebox[\widthof{-1.000}+\arraycolsep][c]{\phantom{-}0.451} & \mathmakebox[\widthof{-1.000}+\arraycolsep][c]{\phantom{-}0.122} & \mathmakebox[\widthof{-1.000}+\arraycolsep][c]{\phantom{-}0} & \mathmakebox[\widthof{-1.000}+\arraycolsep][c]{\phantom{-}0} & \mathmakebox[\widthof{-1.000}+\arraycolsep][c]{\phantom{-}0.064} \\ 
  \mathmakebox[\widthof{-1.000}+\arraycolsep][c]{\phantom{-}0.491} & \mathmakebox[\widthof{-1.000}+\arraycolsep][c]{\phantom{-}0.351} & \mathmakebox[\widthof{-1.000}+\arraycolsep][c]{\phantom{-}0} & \mathmakebox[\widthof{-1.000}+\arraycolsep][c]{\phantom{-}0.158} & \mathmakebox[\widthof{-1.000}+\arraycolsep][c]{\phantom{-}0} & \mathmakebox[\widthof{-1.000}+\arraycolsep][c]{\phantom{-}0} & \mathmakebox[\widthof{-1.000}+\arraycolsep][c]{\phantom{-}0} \\ 
  \mathmakebox[\widthof{-1.000}+\arraycolsep][c]{\phantom{-}0.248} & \mathmakebox[\widthof{-1.000}+\arraycolsep][c]{\phantom{-}0.157} & \mathmakebox[\widthof{-1.000}+\arraycolsep][c]{\phantom{-}0.262} & \mathmakebox[\widthof{-1.000}+\arraycolsep][c]{\phantom{-}0} & \mathmakebox[\widthof{-1.000}+\arraycolsep][c]{\phantom{-}0.191} & \mathmakebox[\widthof{-1.000}+\arraycolsep][c]{\phantom{-}0.142} & \mathmakebox[\widthof{-1.000}+\arraycolsep][c]{\phantom{-}0} \\ 
  \mathmakebox[\widthof{-1.000}+\arraycolsep][c]{\phantom{-}0.303} & \mathmakebox[\widthof{-1.000}+\arraycolsep][c]{\phantom{-}0} & \mathmakebox[\widthof{-1.000}+\arraycolsep][c]{\phantom{-}0} & \mathmakebox[\widthof{-1.000}+\arraycolsep][c]{\phantom{-}0.144} & \mathmakebox[\widthof{-1.000}+\arraycolsep][c]{\phantom{-}0} & \mathmakebox[\widthof{-1.000}+\arraycolsep][c]{\phantom{-}0.553} & \mathmakebox[\widthof{-1.000}+\arraycolsep][c]{\phantom{-}0} \\ 
  \mathmakebox[\widthof{-1.000}+\arraycolsep][c]{\phantom{-}0.214} & \mathmakebox[\widthof{-1.000}+\arraycolsep][c]{\phantom{-}0} & \mathmakebox[\widthof{-1.000}+\arraycolsep][c]{\phantom{-}0} & \mathmakebox[\widthof{-1.000}+\arraycolsep][c]{\phantom{-}0.097} & \mathmakebox[\widthof{-1.000}+\arraycolsep][c]{\phantom{-}0.496} & \mathmakebox[\widthof{-1.000}+\arraycolsep][c]{\phantom{-}0} & \mathmakebox[\widthof{-1.000}+\arraycolsep][c]{\phantom{-}0.194} \\ 
  \mathmakebox[\widthof{-1.000}+\arraycolsep][c]{\phantom{-}0} & \mathmakebox[\widthof{-1.000}+\arraycolsep][c]{\phantom{-}0.224} & \mathmakebox[\widthof{-1.000}+\arraycolsep][c]{\phantom{-}0} & \mathmakebox[\widthof{-1.000}+\arraycolsep][c]{\phantom{-}0} & \mathmakebox[\widthof{-1.000}+\arraycolsep][c]{\phantom{-}0} & \mathmakebox[\widthof{-1.000}+\arraycolsep][c]{\phantom{-}0.776} & \mathmakebox[\widthof{-1.000}+\arraycolsep][c]{\phantom{-}0}

                \end{pmatrix},
\end{align}
where both rows and columns represent treatment nodes. 
Each element $T_{jk}$ describes the probability of a walker currently at node $v_j$, hopping to node $v_k$ in the next step. 
For example, the third row of $\bm{T}$ describes the transition probabilities for a walker at node $v_3:$ IXE\_Q4W. 
From here, the walker can hop to adjacent nodes; $v_1:$ ETN with probability 0.491, $v_2:$ IXE\_Q2W with probability 0.351, and $v_4:$ PBO with probability 0.158. 
The probability of hopping to any other node (not connected to IXE\_Q4W) is zero. 
As required, these probabilities sum to one.\footnotemark[4]\footnotetext[4]{NB: All rows of the transition matrices sum to exactly one, although this may not always appear to be the case due to rounding. } 

\subsection{Bipartite evidence flow}
\subsubsection{Arm-level hat matrix}
Next, we apply the arm-level model to the psoriasis dataset and, for the same comparison of interest, explore the evidence flow on the bipartite graph.
The arm-level hat matrix $\bm{H}_{\top\bot}$ for the psoriasis dataset is shown in Section \ref{app:psoriasis} of the Supplement along with the trial-level hat matrix $\bm{H}$ and their component matrices $\bm{X}$, $\bm{C}$, and $\bm{\Sigma}$ ($=\bm{W}^{-1}$). 
Each row of $\bm{H}_{\top\bot}$ corresponds to a comparison with the baseline treatment and 
each column represents an edge in the bipartite graph (Figure \ref{fig:psoriasis}~(B)). 
These columns are ordered according to the convention $([u_1,v_2], \hdots, [u_1,v_N], [u_2,v_1],\hdots,[u_2,v_N],\hdots,\allowbreak[u_M,v_1],\allowbreak\hdots, [u_M,v_N])$. 
If trial $i$ does not involve treatment $j$, then there is no edge $[u_i,v_j]$ and this column does not appear in $\bm{H}_{\top\bot}$.

As above, we obtain the hat matrix elements for the comparison of interest by applying the consistency equations,
\begin{align*} \bm{H}_{\top\bot}^{(\text{IXE\_Q2W, SEC\_300})}&= \bm{H}_{\top\bot}^{(\text{ETN, SEC\_300})}-\bm{H}_{\top\bot}^{(\text{ETN, IXE\_Q2W})} \\ &\hspace{-45pt}=(0.127, -0.127, -0.118, -0.096, 0.214, -0.006, -0.043, 0.049, -0.621, -0.104, 0.157, 0.567,-0.127,  0.127,\\ &\hspace{-32pt} -0.025, -0.019, 0.043, -0.265, 0.109, 0.156, 0.303, -0.267, -0.056, 0.021, 0.318, -0.340, -0.053, 0.076). 
\end{align*}
Panel (B) in Figure \ref{fig:psoriasis-flow} shows the corresponding evidence flow network from IXE\_Q2W to SEC\_300 on the bipartite graph. 
As before, we observe that this comparison relies strongly on indirect evidence via treatment ETN. 
More specifically, we now also observe that this indirect evidence arises primarily from the UNCOVER-2, UNCOVER-3 and FIXTURE trials. 
Each of these trials also involves a comparison to placebo, but the evidence flow is much smaller through these arms compared with the active intervention arms.
On inspection of the data, we discover that placebo arms have lower event rates which causes them to have lower precision and therefore, less influence.

\subsubsection{Random walk on the bipartite graph}
Next, we define transition probabilities for the random walk on the bipartite graph in Figure \ref{fig:psoriasis}~(B). 
Each edge represents the arm of a trial and is weighted by the inverse of the variance associated with the observation in that arm. 
We use these weights to construct the weighted biadjacency matrix $\bm{B}$ shown in Appendix \ref{app:psoriasis}. 
Via Equation (\ref{eq:Pdown}), we then obtain the transition matrix of the downward walk,
\begin{align}\label{eq:Pdown-psor}
\bm{P}^{\downarrow}=\begin{pmatrix}% latex table generated in R 4.4.1 by xtable 1.8-4 package
% Wed Feb 26 16:14:26 2025
 \mathmakebox[\widthof{-1.000}+\arraycolsep][c]{\phantom{-}0} & \mathmakebox[\widthof{-1.000}+\arraycolsep][c]{\phantom{-}0} & \mathmakebox[\widthof{-1.000}+\arraycolsep][c]{\phantom{-}0} & \mathmakebox[\widthof{-1.000}+\arraycolsep][c]{\phantom{-}0} & \mathmakebox[\widthof{-1.000}+\arraycolsep][c]{\phantom{-}0} & \mathmakebox[\widthof{-1.000}+\arraycolsep][c]{\phantom{-}0.330} & \mathmakebox[\widthof{-1.000}+\arraycolsep][c]{\phantom{-}0.670} \\ 
  \mathmakebox[\widthof{-1.000}+\arraycolsep][c]{\phantom{-}0} & \mathmakebox[\widthof{-1.000}+\arraycolsep][c]{\phantom{-}0} & \mathmakebox[\widthof{-1.000}+\arraycolsep][c]{\phantom{-}0} & \mathmakebox[\widthof{-1.000}+\arraycolsep][c]{\phantom{-}0.113} & \mathmakebox[\widthof{-1.000}+\arraycolsep][c]{\phantom{-}0.509} & \mathmakebox[\widthof{-1.000}+\arraycolsep][c]{\phantom{-}0.378} & \mathmakebox[\widthof{-1.000}+\arraycolsep][c]{\phantom{-}0} \\ 
  \mathmakebox[\widthof{-1.000}+\arraycolsep][c]{\phantom{-}0} & \mathmakebox[\widthof{-1.000}+\arraycolsep][c]{\phantom{-}0} & \mathmakebox[\widthof{-1.000}+\arraycolsep][c]{\phantom{-}0} & \mathmakebox[\widthof{-1.000}+\arraycolsep][c]{\phantom{-}0.020} & \mathmakebox[\widthof{-1.000}+\arraycolsep][c]{\phantom{-}0.517} & \mathmakebox[\widthof{-1.000}+\arraycolsep][c]{\phantom{-}0.462} & \mathmakebox[\widthof{-1.000}+\arraycolsep][c]{\phantom{-}0} \\ 
  \mathmakebox[\widthof{-1.000}+\arraycolsep][c]{\phantom{-}0.355} & \mathmakebox[\widthof{-1.000}+\arraycolsep][c]{\phantom{-}0} & \mathmakebox[\widthof{-1.000}+\arraycolsep][c]{\phantom{-}0} & \mathmakebox[\widthof{-1.000}+\arraycolsep][c]{\phantom{-}0.070} & \mathmakebox[\widthof{-1.000}+\arraycolsep][c]{\phantom{-}0.322} & \mathmakebox[\widthof{-1.000}+\arraycolsep][c]{\phantom{-}0.254} & \mathmakebox[\widthof{-1.000}+\arraycolsep][c]{\phantom{-}0} \\ 
  \mathmakebox[\widthof{-1.000}+\arraycolsep][c]{\phantom{-}0} & \mathmakebox[\widthof{-1.000}+\arraycolsep][c]{\phantom{-}0.197} & \mathmakebox[\widthof{-1.000}+\arraycolsep][c]{\phantom{-}0} & \mathmakebox[\widthof{-1.000}+\arraycolsep][c]{\phantom{-}0} & \mathmakebox[\widthof{-1.000}+\arraycolsep][c]{\phantom{-}0} & \mathmakebox[\widthof{-1.000}+\arraycolsep][c]{\phantom{-}0} & \mathmakebox[\widthof{-1.000}+\arraycolsep][c]{\phantom{-}0.803} \\ 
  \mathmakebox[\widthof{-1.000}+\arraycolsep][c]{\phantom{-}0} & \mathmakebox[\widthof{-1.000}+\arraycolsep][c]{\phantom{-}0} & \mathmakebox[\widthof{-1.000}+\arraycolsep][c]{\phantom{-}0} & \mathmakebox[\widthof{-1.000}+\arraycolsep][c]{\phantom{-}0.111} & \mathmakebox[\widthof{-1.000}+\arraycolsep][c]{\phantom{-}0.567} & \mathmakebox[\widthof{-1.000}+\arraycolsep][c]{\phantom{-}0.322} & \mathmakebox[\widthof{-1.000}+\arraycolsep][c]{\phantom{-}0} \\ 
  \mathmakebox[\widthof{-1.000}+\arraycolsep][c]{\phantom{-}0} & \mathmakebox[\widthof{-1.000}+\arraycolsep][c]{\phantom{-}0.300} & \mathmakebox[\widthof{-1.000}+\arraycolsep][c]{\phantom{-}0.527} & \mathmakebox[\widthof{-1.000}+\arraycolsep][c]{\phantom{-}0.172} & \mathmakebox[\widthof{-1.000}+\arraycolsep][c]{\phantom{-}0} & \mathmakebox[\widthof{-1.000}+\arraycolsep][c]{\phantom{-}0} & \mathmakebox[\widthof{-1.000}+\arraycolsep][c]{\phantom{-}0} \\ 
  \mathmakebox[\widthof{-1.000}+\arraycolsep][c]{\phantom{-}0.481} & \mathmakebox[\widthof{-1.000}+\arraycolsep][c]{\phantom{-}0.154} & \mathmakebox[\widthof{-1.000}+\arraycolsep][c]{\phantom{-}0.340} & \mathmakebox[\widthof{-1.000}+\arraycolsep][c]{\phantom{-}0.025} & \mathmakebox[\widthof{-1.000}+\arraycolsep][c]{\phantom{-}0} & \mathmakebox[\widthof{-1.000}+\arraycolsep][c]{\phantom{-}0} & \mathmakebox[\widthof{-1.000}+\arraycolsep][c]{\phantom{-}0} \\ 
  \mathmakebox[\widthof{-1.000}+\arraycolsep][c]{\phantom{-}0.491} & \mathmakebox[\widthof{-1.000}+\arraycolsep][c]{\phantom{-}0.178} & \mathmakebox[\widthof{-1.000}+\arraycolsep][c]{\phantom{-}0.244} & \mathmakebox[\widthof{-1.000}+\arraycolsep][c]{\phantom{-}0.087} & \mathmakebox[\widthof{-1.000}+\arraycolsep][c]{\phantom{-}0} & \mathmakebox[\widthof{-1.000}+\arraycolsep][c]{\phantom{-}0} & \mathmakebox[\widthof{-1.000}+\arraycolsep][c]{\phantom{-}0} 
  
\end{pmatrix},
\end{align}
where rows represent trials and columns represent treatments. 
Each element $P^{\downarrow}_{ij}$ describes the probability of a walker currently at trial node $u_i$, hopping to treatment node $v_j$ in the next step. 
For example, the first row contains the transition probabilities for a walker at trial node $u_1:$ CLEAR. 
This trial has two arms, and the walker can hop to either one of these treatments in the next step; $v_6:$ SEC\_300 with probability 0.330, or $v_7:$ UST with probability 0.670. 
The probability of hopping to another trial node, or to a treatment that does not appear in that trial is zero.

Similarly, we obtain the upstream transition matrix via Equation (\ref{eq:Pup}),
\begin{align}\label{eq:Pup-psor}
    \bm{P}^{\uparrow}=\begin{pmatrix}% latex table generated in R 4.4.1 by xtable 1.8-4 package
% Wed Feb 26 16:14:26 2025
 \mathmakebox[\widthof{-1.000}+\arraycolsep][c]{\phantom{-}0} & \mathmakebox[\widthof{-1.000}+\arraycolsep][c]{\phantom{-}0} & \mathmakebox[\widthof{-1.000}+\arraycolsep][c]{\phantom{-}0} & \mathmakebox[\widthof{-1.000}+\arraycolsep][c]{\phantom{-}0.307} & \mathmakebox[\widthof{-1.000}+\arraycolsep][c]{\phantom{-}0} & \mathmakebox[\widthof{-1.000}+\arraycolsep][c]{\phantom{-}0} & \mathmakebox[\widthof{-1.000}+\arraycolsep][c]{\phantom{-}0} & \mathmakebox[\widthof{-1.000}+\arraycolsep][c]{\phantom{-}0.329} & \mathmakebox[\widthof{-1.000}+\arraycolsep][c]{\phantom{-}0.364} \\ 
  \mathmakebox[\widthof{-1.000}+\arraycolsep][c]{\phantom{-}0} & \mathmakebox[\widthof{-1.000}+\arraycolsep][c]{\phantom{-}0} & \mathmakebox[\widthof{-1.000}+\arraycolsep][c]{\phantom{-}0} & \mathmakebox[\widthof{-1.000}+\arraycolsep][c]{\phantom{-}0} & \mathmakebox[\widthof{-1.000}+\arraycolsep][c]{\phantom{-}0.062} & \mathmakebox[\widthof{-1.000}+\arraycolsep][c]{\phantom{-}0} & \mathmakebox[\widthof{-1.000}+\arraycolsep][c]{\phantom{-}0.352} & \mathmakebox[\widthof{-1.000}+\arraycolsep][c]{\phantom{-}0.260} & \mathmakebox[\widthof{-1.000}+\arraycolsep][c]{\phantom{-}0.325} \\ 
  \mathmakebox[\widthof{-1.000}+\arraycolsep][c]{\phantom{-}0} & \mathmakebox[\widthof{-1.000}+\arraycolsep][c]{\phantom{-}0} & \mathmakebox[\widthof{-1.000}+\arraycolsep][c]{\phantom{-}0} & \mathmakebox[\widthof{-1.000}+\arraycolsep][c]{\phantom{-}0} & \mathmakebox[\widthof{-1.000}+\arraycolsep][c]{\phantom{-}0} & \mathmakebox[\widthof{-1.000}+\arraycolsep][c]{\phantom{-}0} & \mathmakebox[\widthof{-1.000}+\arraycolsep][c]{\phantom{-}0.377} & \mathmakebox[\widthof{-1.000}+\arraycolsep][c]{\phantom{-}0.350} & \mathmakebox[\widthof{-1.000}+\arraycolsep][c]{\phantom{-}0.273} \\ 
  \mathmakebox[\widthof{-1.000}+\arraycolsep][c]{\phantom{-}0} & \mathmakebox[\widthof{-1.000}+\arraycolsep][c]{\phantom{-}0.152} & \mathmakebox[\widthof{-1.000}+\arraycolsep][c]{\phantom{-}0.007} & \mathmakebox[\widthof{-1.000}+\arraycolsep][c]{\phantom{-}0.218} & \mathmakebox[\widthof{-1.000}+\arraycolsep][c]{\phantom{-}0} & \mathmakebox[\widthof{-1.000}+\arraycolsep][c]{\phantom{-}0.033} & \mathmakebox[\widthof{-1.000}+\arraycolsep][c]{\phantom{-}0.296} & \mathmakebox[\widthof{-1.000}+\arraycolsep][c]{\phantom{-}0.061} & \mathmakebox[\widthof{-1.000}+\arraycolsep][c]{\phantom{-}0.233} \\ 
  \mathmakebox[\widthof{-1.000}+\arraycolsep][c]{\phantom{-}0} & \mathmakebox[\widthof{-1.000}+\arraycolsep][c]{\phantom{-}0.337} & \mathmakebox[\widthof{-1.000}+\arraycolsep][c]{\phantom{-}0.086} & \mathmakebox[\widthof{-1.000}+\arraycolsep][c]{\phantom{-}0.494} & \mathmakebox[\widthof{-1.000}+\arraycolsep][c]{\phantom{-}0} & \mathmakebox[\widthof{-1.000}+\arraycolsep][c]{\phantom{-}0.083} & \mathmakebox[\widthof{-1.000}+\arraycolsep][c]{\phantom{-}0} & \mathmakebox[\widthof{-1.000}+\arraycolsep][c]{\phantom{-}0} & \mathmakebox[\widthof{-1.000}+\arraycolsep][c]{\phantom{-}0} \\ 
  \mathmakebox[\widthof{-1.000}+\arraycolsep][c]{\phantom{-}0.196} & \mathmakebox[\widthof{-1.000}+\arraycolsep][c]{\phantom{-}0.264} & \mathmakebox[\widthof{-1.000}+\arraycolsep][c]{\phantom{-}0.081} & \mathmakebox[\widthof{-1.000}+\arraycolsep][c]{\phantom{-}0.410} & \mathmakebox[\widthof{-1.000}+\arraycolsep][c]{\phantom{-}0} & \mathmakebox[\widthof{-1.000}+\arraycolsep][c]{\phantom{-}0.050} & \mathmakebox[\widthof{-1.000}+\arraycolsep][c]{\phantom{-}0} & \mathmakebox[\widthof{-1.000}+\arraycolsep][c]{\phantom{-}0} & \mathmakebox[\widthof{-1.000}+\arraycolsep][c]{\phantom{-}0} \\ 
  \mathmakebox[\widthof{-1.000}+\arraycolsep][c]{\phantom{-}0.673} & \mathmakebox[\widthof{-1.000}+\arraycolsep][c]{\phantom{-}0} & \mathmakebox[\widthof{-1.000}+\arraycolsep][c]{\phantom{-}0} & \mathmakebox[\widthof{-1.000}+\arraycolsep][c]{\phantom{-}0} & \mathmakebox[\widthof{-1.000}+\arraycolsep][c]{\phantom{-}0.327} & \mathmakebox[\widthof{-1.000}+\arraycolsep][c]{\phantom{-}0} & \mathmakebox[\widthof{-1.000}+\arraycolsep][c]{\phantom{-}0} & \mathmakebox[\widthof{-1.000}+\arraycolsep][c]{\phantom{-}0} & \mathmakebox[\widthof{-1.000}+\arraycolsep][c]{\phantom{-}0} 
  
\end{pmatrix}.
\end{align}
Here, rows represent treatments and columns represent trials, such that an element $P^{\uparrow}_{ji}$ describes the probability of a walker currently at bottom node $v_j$, hopping to top node $u_i$ in the next step. 
Similar to matrix $\bm{T}$ in Equation (\ref{eq:T-psor}), the third row of $\bm{P}^{\uparrow}$ contains the transition probabilities for a walker at treatment node $v_3:$ IXE\_Q4W. 
However, rather than hopping to another treatment node, this walker can only hop to the trials (top nodes) in which it appears; $u_7:$ UNCOVER-1 with probability 0.377, $u_8:$ UNCOVER-2 with probability 0.350 and $u_9:$ UNCOVER-3 with probability 0.273. 
The probability of hopping to another treatment node, or to a trial that does not contain IXE\_Q4W is zero. 

\subsection{Conjectures}
Finally, we use our results for the psoriasis dataset to check whether our two conjectures hold. 

\subsubsection{Conjecture 1}
To check Conjecture 1, we compare the arm-level hat matrix $\bm{H}_{\top\bot}$ with the matrix of nodal currents $\bm{I}'$ calculated via Equation (\ref{eq:Iprime}). 
In Section \ref{app:psoriasis} of the Supplementary Material we show $\bm{I}'$ for the psoriasis data along with its component matrices $\bm{J}'$, $\bm{B}_{\top\bot}$, and $\bm{R}$. 
Comparing each element of $\boldsymbol{I}'$ with the corresponding element of $\bm{H}_{\top\bot}$ gives a maximum absolute difference of $2.44\times10^{-15}$. 
As a general rule of thumb, two values can be considered numerically identical if their absolute difference is less than the square root of the machine precision (\cite{Higham:2002}). 
For the machine used in these calculations, this value was $1.49\times10^{-8}$. 
Therefore, for the psoriasis dataset, these two matrices are numerically identical, supporting Conjecture 1. 
%Max difference 2.442491e-15

\subsubsection{Conjecture 2}
To check Conjecture 2, we calculate the two-step transition matrix $\boldsymbol{P}$ via Equation (\ref{eq:P}) using the upstream and downstream transition matrices in Equations (\ref{eq:Pdown-psor}) and (\ref{eq:Pup-psor}). 
Each element $P_{jk}$ of this matrix describes the probability that a walker currently at bottom node $v_j$ on the bipartite graph hops to bottom node $v_k$ in the next two consecutive steps (via any top node). 
The renormalised transition matrix $\bm{\tilde{P}}$ is obtained by setting the diagonal elements of $\bm{P}$ equal to zero and renormalising the remaining probabilities in each row. In Section \ref{app:psoriasis} of the Supplement we show both $\bm{P}$ and $\bm{\tilde{P}}$ for the psoriasis dataset. 
Comparing each element of $\bm{\tilde{P}}$ with the corresponding element of $\bm{T}$ in Equation (\ref{eq:T-psor}) gives a maximum absolute difference of $1.33\times10^{-15}$.
Therefore, these two matrices can be considered numerically identical, meaning Conjecture 2 holds for this dataset.

%Max difference 1.332268e-15

\section{Simulations}\label{sec:simulations}
In Section \ref{sec:application} we showed that Conjectures 1 and 2 hold true for the psoriasis dataset. 
To investigate this more thoroughly, we  evaluate the two conjectures for 10,000 randomly generated NMA graphs. We describe our simulation method in Section \ref{app:simulation} of the Supplementary Material. 
This involves generating random network structures (i.e. which treatments appear in which trials) as well as sampling the outcome data in each trial. 
As shown in the Supplement, the bipartite representation of NMA provides a natural framework for automatically generating networks with multi-arm trials.

For each randomly generated NMA, we evaluate Conjecture 1 by comparing each element of $\bm{H}_{\top\bot}$ with the corresponding element of $\bm{I}'$. Across all 10,000 networks, the maximum difference between two equivalent elements was $6.03 \times 10^{-11}$. Similarly, to evaluate Conjecture 2 we compare $\bm{T}$ and $\bm{\tilde{P}}$ for each network. Here, the maximum difference between two corresponding elements was $1.15\times 10^{-13}$. This provides strong empirical evidence for both conjectures.   

\section{Summary and discussion}\label{sec:discuss}

In this paper, we have formalised the bipartite framework for NMA.
Treatments and trials define two distinct types of node, while edges correspond to the arms of a trial, connecting each trial node to the treatments it compares. 
This representation accurately describes the data structure of NMA, showing which treatments have been compared in which trials. 
Unlike the traditional representation of NMA as a unipartite graph, the bipartite graph shows comparisons from trials that compare more than two treatments.

Understanding how evidence from different trials combines to give overall estimates of treatment effects is a key challenge for NMA. 
The influence of observations from the different trials is related to the structure of the network and the weight assigned to each observation in the model. 
This information is captured in the model's hat matrix. 
In earlier work, \cite{Konig:2013} used the hat matrix of the aggregate model to define an evidence flow network on the unipartite NMA graph.
However, it is not clear how these flows relate to the contributions of individual trials.
Our work overcomes this limitation by instead characterizing the evidence flow on the bipartite graph.
We considered an `arm level' parameterization of the standard NMA model that expresses the treatment effect estimates as a linear combination of the observations in each arm of each trial.
By linking this model to the bipartite framework, we showed that the elements in each row of the arm-level hat matrix define the evidence flow through each edge in the bipartite graph.
This approach reveals how evidence flows through the individual trials.

In practice, arm-level observations will not always be available. 
Instead, trials might report relative effects with respect to some reference arm.
However, to construct the arm-level hat matrix we only need the variances associated with the observations.
In fact, the arm-level hat matrix is just a projection of the standard trial-level hat matrix which depends on the variances and covariances of the contrast-level observations.
The covariance between contrast-level observations in a trial is the variance in the reference arm of that trial.
If this information is not available, it can be approximated (see e.g. \cite{TSD2, Riley:2009}).
Assuming arms are independent, the arm-level variances can be reconstructed from the contrast-level variances and the variance in the reference arm.
Therefore, the arm-level hat matrix requires no additional information compared with the standard model.

Using the analogy between random walks and evidence flow set out in \cite{Davies:2022}, we defined a higher order random walk on the bipartite graph.
The walk is made up of two parts: `upward' transitions from treatments to trials, and `downward' transitions from trials to treatments. 
Previously, \cite{Davies:2022} showed that the movement of a random walker on the NMA graph contains information about the propagation of evidence through the network.
Therefore, we proposed a conjecture linking the HoRW to the bipartite evidence flow network.
This provides a random walk interpretation for the flow of evidence through the arms of the trials.
We then proposed a second conjecture, describing how the original evidence flows on the unipartite graph can be derived from the HoRW. 
This result reveals the connection between the evidence structures in the unipartite and bipartite frameworks.
We verified both conjectures in an application to a real dataset and in extensive simulations on randomly generated graphs. 
However, theoretical proofs of these conjectures remain an open challenge.

One limitation of the bipartite framework as a visualization of NMA is that these graphs quickly become large and complex as the number of treatments and trials increases. 
For larger networks, it is increasingly difficult to arrange the bipartite graph in a way that clearly shows the flow of evidence between two treatments.
This points to a need for more sophisticated visualization tools developed specifically for this framework. 
We note, however, that the utility of the bipartite framework for NMA is not only related to visualization. 
The bipartite graph is an accurate mathematical description of the data structure.
By setting out this framework, NMA can access the suite of tools and techniques developed for the study of bipartite structures in other areas of network science.

Indeed, the bipartite framework for NMA opens up many avenues for future applications and developments.
A natural next step of the work we have described is to use the flow of information in the bipartite graph to evaluate the influence of trials. 
This is the subject of ongoing research that we expect will overcome the limitations of previous attempts at this task (e.g \cite{Lu:2011, Krahn:2013, Jackson:2017, Riley:2018, Copas:2018, Papakon:2018, Rucker:2020}). 
We also predict that the random walk interpretation of evidence flow will provide a numerical tool to uncover the contribution of trials in more complex multi-parameter evidence synthesis models. 
In our simulations, we made use of a simple bipartite random graph model to generate synthetic networks with multi-arm trials. 
In another parallel project, we are further developing this algorithm to produce networks with more realistic characteristics.
Related to this, previous work has attempted to characterize the structure (or geometry) of NMA networks (e.g. \cite{Salanti:2008b, Nikola:2014, Tonin:2019, Papakon:2020}).
Using the bipartite description of NMA, we can more accurately evaluate the properties of these data structures.
It will then be interesting to investigate how features of the bipartite graph are related to the performance of the NMA.
Finally, \cite{Lumley:2024} proposed the bipartite representation of NMA as a tool to identify inconsistencies between different sources of evidence in the network.
He predicted that this would alleviate the challenges faced when loops of evidence involve estimates from multi-arm trials.

This work marks the first application of higher-order network theory in the context of NMA. 
The bipartite framework provides new insights into the evidence structure and the role of individual trials in NMA estimates. 
It also lays the groundwork for future methodological advancements and applications.
The study of higher order interactions is a young and actively evolving research area in network science. 
By framing NMA in this context, we have enabled access to state of the art techniques developed in this area.
Moreover, there is now an exciting opportunity for NMA to shape the direction of research in this emerging field.

% Acknowledgments ------------------------------------------
\section*{Acknowledgments}
ALD acknowledges funding from the Engineering and Physical Sciences Research Council (EP/Y007905/1). 
I would like to thank my fellowship advisory group for their valuable insights and feedback throughout the progression of this paper:  Gerta R{\"u}cker, Ian White, Nicky Welton, Julian Higgins, Ayavaldi Ganesh, and Guido Schwarzer.
I also thank David Phillippo for many useful discussions particularly regarding notation and presentation. 
\vspace{40pt}

% Bibliography ------------------------------------------
\bibliographystyle{abbrvnat}
\bibliography{refs.bib}

% Appendices --------------------------------------------
\clearpage
\begin{appendices}
% Turn on appendix numbering
%\setcounter{secnumdepth}{3}

\setcounter{page}{1}
\setcounter{section}{0}
\setcounter{equation}{0}
\setcounter{table}{0}
\renewcommand{\theequation}{\Alph{section}.\arabic{equation}}
\renewcommand{\thetable}{\Alph{section}.\arabic{table}}
\renewcommand{\thefigure}{\Alph{section}.\arabic{figure}}

\begin{center}
\section*{Supplementary Material}
\end{center}
\section{Design matrix} \label{app:design}
The design matrix in the definition of the trial-level hat matrix (Equation (\ref{eq:model}) in the main text) describes which treatments are compared in each trial. Each trial contributes an $(n_i-1) \times (N-1)$ matrix $\bm{X_i}$ such that $\bm{X} = [X_1 \hspace{5pt} X_2 \hspace{5pt} \hdots \hspace{5pt} X_M]'$. 
Each of the $N-1$ columns of $\bm{X}_i$ represents a comparison between treatment $j=2,\hdots,N$ and the global baseline $j=1$. The $n_i-1$ rows represent the comparisons of treatments $[t_{i,\ell};\ell=2,\hdots, n_i]$ to the trial-specific baseline $t_{i,1}$. A given row $\ell$ contains a $+1$ in the column corresponding to $j=t_{i,\ell}$. If the trial-specific baseline is not the global baseline ($t_{i,1}\neq 1$), then there is a $-1$ in the column corresponding to $j=t_{i,1}$. All other entries are zero. 

\section{Aggregate Model}\label{app:agg}
In this section, we summarize the aggregate version of the graph theoretical model.
For details of this model, and how it corresponds to other formulations we refer to \cite{Davies:2022} and the Supplementary Materials therein. 

Each $n_i$-armed trial is associated with $q_i=n_i(n_i-1)/2$ contrast-level observations representing comparisons between each pair of treatments in the trial.  
For a trial with $n_i=2$ arms, the weight of each observation is given by its inverse variance $w_{i,t_{i\ell}t_{i\ell'}}=(\sigma_{t_{i\ell}t_{i\ell'}}^2+\tau^2)^{-1}$ (where $\tau=0$ for a CE model and must be estimated in an RE model). For a trial with $n_i>2$, we account for correlations by adjusting the variances using a method described in \cite{Rucker:2012, Rucker:2014} and \cite{Gutman:2004}. This results in a set of $q_i$ adjusted weights $[w_{i,t_{i\ell}t_{i\ell'}}; \ell=1,\hdots,n_i-1, \ell'=\ell+1,\hdots,n_i]$ describing a fully-connected subgraph of $q_i$ 2-armed trials that is equivalent to the original multi-arm trial. 

In step 1 of the aggregate model, we use the adjusted weights to perform a pairwise meta-analysis for each pair of treatments $jk$ that have been compared in at least one trial (i.e. across each of the $K_\bot$ edges in the unipartite graph). 
The resulting direct estimate of this comparison is a weighted mean
\begin{align}
    \hat{\theta}_{jk}^{\text{dir}} = \frac{\sum_{i\in \mathcal{M}_{jk} }w_{i,jk}y_{i,jk}}{\sum_{i\in \mathcal{M}_{jk} }w_{i,jk}},
\end{align}
where we have written $\mathcal{M}_{jk}$ for the set of trials involving treatments $j$ and $k$. That is, $\mathcal{M}_{jk}=[i|\mathcal{T}_i\supseteq [j,k]]$. The inverse variance of the direct estimate is then
\begin{align}
    w_{jk} = \sum_{i\in \mathcal{M}_{jk} }w_{i,jk}.
\end{align}
We collect the direct estimates in a  $K_\bot$-dimensional vector $\bm{\hat\theta^{\text{dir}}}$ and the inverse-variances in a $K_\bot \times K_\bot$ diagonal matrix $\bm{W_\bot}=\text{diag}(w_{jk})$.
The matrix $\bm{W}_\bot$ then defines a set of edge weights, where each element $w_{jk}$ is the weight associated with edge $[v_j,v_k]$ in the unipartite graph.

In the second step, we use the edge weights to combine the direct estimates in an NMA. To do so, we make use of the oriented edge-vertex incidence matrix of the unipartite graph $\bm{B_\bot}$ described in Section \ref{sec:simple} in the main paper. Without loss of generality, each row representing an edge $[v_j,v_k]$ is given an arbitrary orientation, with a $-1$ in column $j$ and $+1$ in column $k$. The network estimates of the basic parameters are then obtained via 
\begin{align}
    \bm{\hat\theta} = \bm{H_\bot}\bm{\hat\theta^{\text{dir}}},
\end{align}
where we write $\bm{H_\bot}$ for the $(N-1)\times K_\bot$ aggregate hat matrix, defined as
\begin{align}\label{eq:H_bot2}
    \bm{H_\bot} = \bm{C}_N (\bm{B'_\bot}\bm{W_\bot}\bm{B_\bot})^{+}\bm{B'_\bot}\bm{W_\bot}.
\end{align}
The $N\times N$ matrix $\bm{L_\bot}=\bm{B'_\bot}\bm{W_\bot}\bm{B_\bot}$ is the Laplacian of the weighted unipartite graph and $\bm{L^+_\bot}$ denotes its Moore-Penrose pseudo-inverse (\cite{Gutman:2004}). 
The Laplacian is a matrix representation of a graph, related to many of its key properties. 
For more information, we refer the interested reader to various sources e.g. \cite{Rucker:2012, Gutman:2004, Godsil:2001}, and \cite{Mohar:2005}. 
Recall from the main paper (Section \ref{sec:setup}) the definition of matrix $\bm{C}_{N}=\left[ -\bm{1}_{N-1} \hspace{5pt} \bm{I}_{N-1} \right]$. 
In Equation (\ref{eq:H_bot2}), this matrix ensures that $\bm{H_\bot}$ has the correct dimensions for projecting onto the $N-1$ basic parameters. 
Similar to the trial-level hat matrix, each row of $\bm{H_\bot}$ contains the coefficients of a linear equation describing how the estimates of each basic parameter $\hat\theta_{1j}$ depend on the direct estimates associated with each edge in the unipartite graph, $\hat\theta^{\text{dir}}_{jk}; [v_j,v_k] \in E_\bot$. As before, we obtain the full version of the aggregate hat matrix by applying the consistency equations in Equation (\ref{eq:hat-consistency}) in the main paper.

\section{Edge vertex incidence matrix of the bipartite graph}\label{app:bi-incidence}
Figure \ref{fig:HOI-rep}~(D) in the main paper shows the bipartite graph for an NMA of six treatments,  $[a,b,c,d,e,f]$, compared in three trials, $u_1=[b,c]$, $u_2=[c,e,f]$, and $u_3=[a,b,d,e]$. The corresponding edge-vertex incidence matrix of this graph is given by
\begin{align}\label{eq:bi-inc-eg}
\bm{B}_{\top\bot}=
\begin{blockarray}{cccccccccc}
 & u_1 & u_2 & u_3 & \phantom{,}a\phantom{,} & \phantom{,}b\phantom{,} & \phantom{,}c\phantom{,} & \phantom{,}d\phantom{,} & \phantom{,}e\phantom{,} & \phantom{,}f\phantom{,} \\
\begin{block}{c(ccccccccc)}
  \lbrack u_1, b\rbrack & 1 & 0 & 0 & 0 & 1 & 0 & 0 & 0 & 0 \\
  \lbrack u_1, c\rbrack & 1 & 0 & 0 & 0 & 0 & 1 & 0 & 0 & 0\\
  \lbrack u_2, a\rbrack & 0 & 1 & 0 & 1 & 0 & 0 & 0 & 0 & 0\\
  \lbrack u_2, b\rbrack & 0 & 1 & 0 & 0 & 1 & 0 & 0 & 0 & 0\\
  \lbrack u_2, d\rbrack & 0 & 1 & 0 & 0 & 0 & 0 & 1 & 0 & 0\\
  \lbrack u_2, e\rbrack & 0 & 1 & 0 & 0 & 0 & 0 & 0 & 1 & 0\\
  \lbrack u_3, c\rbrack & 0 & 0 & 1 & 0 & 0 & 1 & 0 & 0 & 0\\
  \lbrack u_3, e\rbrack & 0 & 0 & 1 & 0 & 0 & 0 & 0 & 1 & 0\\
  \lbrack u_3, f\rbrack & 0 & 0 & 1 & 0 & 0 & 0 & 0 & 0 & 1\\
\end{block}
\end{blockarray}
\hspace{2pt},
\end{align}
where each row represents an edge $\in E_{\top\bot}$ between a top and bottom node, and each column represents a node. 
The first $M$ columns correspond to top nodes $\in V_{\top}$ and the last $N$ columns correspond to bottom nodes $\in V_\bot$.

\section{Electrical networks and bipartite NMA}\label{app:electric}
\subsection{Unipartite framework}
\subsubsection{NMA and electrical networks}
In an electrical network, edges represent resistors that connect at the nodes. If two (or more) nodes of an electric network are connected to the poles of a battery, this results in voltages across all edges. The voltages in turn induce currents across the edges (current = voltage divided by resistance). Currents may also flow into or out of a node from or to the external battery (often referred to simply as the “exterior”). 

\cite{Rucker:2012}'s analogy between NMA and electrical networks   is based on the observation that resistors in parallel and sequential circuits combine in the same way as variance in pairwise and indirect meta-analysis. Variance therefore corresponds to resistance. R{\"u}cker used this connection to show that the graph theoretical methods used to derive voltages across edges in an electrical network can be used to calculate network estimates of relative treatment effects in NMA. 

In \cite{Davies:2022} we extended this analogy and proved that the aggregate hat matrix has an interpretation in terms of electrical current.  More precisely, the elements in the row of the hat matrix corresponding to the comparison of treatments $j$ and $k$ can be obtained as follows: Connect a battery to nodes $v_j$ and $v_k$ in the electrical circuit so that one unit of current flows from the exterior into node $v_j$, and out of the network (to the exterior) from node $v_k$. The external currents into/out of all other nodes are zero. This set-up induces currents across the edges in the network. The analogy is then as follows: the current along edge $[v_l, v_m]$ is identical to the hat matrix element $H_{\bot,lm}^{(jk)}$. Further details, examples and a mathematical proof can be found in \cite{Davies:2022}. 

The proof uses Ohm's law and Kirchoff's law to write the currents flowing through each edge in terms of the currents at the nodes. We define a matrix of edge currents $\bm{I}_{\bot}$ and a matrix of nodal currents $\bm{J}_{\bot}$ such that each column represents a different placement of the battery. That is, columns $\bm{I}_{\bot,jk}$ and $\bm{J}_{\bot,jk}$ represent the scenario in which a battery is attached across nodes $v_j$ and $v_k$ (with one unit of current flowing into $v_j$ and out of $v_k$). The element $I_{\bot,jk}^{(lm)}$ is then the current flowing along edge $[v_l, v_m]$ and element $J_{\bot,jk}^{(l)}$ is the current flowing from the exterior to node $v_l$. Nodal currents are defined as follows:
\begin{align}
    J_{\bot,jk}^{(l)} = \begin{cases}
        \phantom{-}1 & \text{if }\hspace{3pt} l=j\\
        -1 & \text{if }\hspace{3pt} l=k\\
        \phantom{-}0 & \text{otherwise}.
    \end{cases}
\end{align}
We specify both matrices $\bm{I}_\bot$ and $\bm{J}_\bot$ with $N-1$ columns such that each placement of the battery is between node $v_1$ one other ($v_j$; $j=2,\hdots,N$). Matrix $\bm{I}_{\bot}$ has $K_\bot$ rows representing each edge in the unipartite graph and $\bm{J}_{\bot}$ has $N$ rows representing each node (treatment). Based on this construction, we find $\bm{J}_\bot = \bm{C}_N'$, where we recall $\bm{C}_{N}=\left[ -\bm{1}_{N-1} \hspace{5pt} \bm{I}_{N-1} \right]$.

The laws of electrical network theory lead to the following equation relating the edge and nodal currents,
\begin{align}
    \bm{I}_\bot = \bm{R}_\bot^{-1} \bm{B}_\bot (\bm{B}_\bot' \bm{R}_\bot^{-1} \bm{B}_\bot)^{+} \bm{J}_\bot,
\end{align}
where $\bm{B}_\bot$ is the (oriented) edge-vertex incidence matrix of the (unipartite) graph and $\bm{R}_\bot$ is a diagonal matrix containing the resistance associated with each edge. Based on R{\"u}cker's analogy, the resistance in each edge of the electrical network is equal to the variance of the direct estimate associated with each edge in the unipartite NMA graph. Therefore, the inverse of the matrix of resistances is equal to the inverse-variance weight matrix of the aggregate model, $\bm{R}_\bot^{-1}=\bm{W}_\bot$. In \cite{Davies:2022} we used this equivalence, along with $\bm{J}_\bot = \bm{C}_N'$ noted above, to obtain the transpose of the matrix of edge currents,
\begin{align}
     \bm{I}_\bot' &= \bm{J}_\bot' ((\bm{B}_\bot' \bm{R}_\bot^{-1} \bm{B}_\bot)^{+})' \bm{B}_\bot ' (\bm{R}_\bot^{-1})'\\
     &= \bm{C}_N (\bm{B}_\bot' \bm{W}_\bot \bm{B}_\bot)^{+} \bm{B}_\bot ' \bm{W}_\bot, \label{eq:I-result}
\end{align}
which is equal to the aggregate hat matrix in Equation (\ref{eq:H_bot}). 

\subsubsection{Random walks and electrical networks}

The relationship between random walks and electrical networks is well known (\cite{DoyleSnell:1984}). For reference, see \cite{Davies:2022} (Section 4.2 and Supplementary Material Section F). We summarize the key points here. 

Starting from an electrical network with resistances $R_{\bot,jk}$ along each edge, a random walk can be constructed by setting the probability of hopping from node $v_j$ to node $v_k$ proportional to the inverse of the resistance in edge $[v_j,v_k]$. Electrical current can then be interpreted in the random-walk picture as follows: When a voltage is applied between two nodes $a$ and $b$ such that the total current flowing into $a$ and out of $b$ from the exterior is 1, the current induced in each edge, $[v_j,v_k]$, is equal to the expected net number of times a random walker, starting at $a$ and walking until it reaches $b$, moves along the edge from $v_j$ to $v_k$. The net number of times the walker moves from $v_j$ to $v_k$ is the number of crossings in the direction from $v_j$ to $v_k$ minus the number of crossings in the opposite direction.

This analogy, along with the result in Equation (\ref{eq:I-result}), is the basis for Theorem 1 in the main paper. 

\subsection{Extension to the bipartite framework}

\subsubsection{NMA and electrical networks}
The electrical network and random walk analogy extends easily to the bipartite framework. 
For an electrical network, the distinction between top and bottom nodes is not important. 
The bipartite graph can be treated simply as a unipartite graph with $N+M$ nodes and $K_{\top\bot}$ edges.
As before, each edge has an associated resistance.
In the NMA context, edges now correspond to the arms of trials meaning the resistance in each edge is equal to the variance associated with each trial arm. 
Therefore, we define an $K_{\top\bot} \times K_{\top\bot}$ diagonal matrix of resistances $\bm{R}$ with diagonal elements equal to $\sigma_{i,j}^2+\frac{\tau^2}{2}$. We retain the notation from the main paper that $i$ indexes trials (top nodes), while $j,k,l,m$ refer to treatments (bottom nodes).

Next, we define the matrix of edge currents, $\bm{I}$, and the matrix of nodal currents, $\bm{J}$, for the bipartite graph. As before, we set the number of columns to $N-1$ representing a battery attached across pairs of treatment nodes, $v_1$ and $v_j; j=2,\hdots,N$. Now, matrix $\bm{I}$ has $K_{\top\bot}$ rows representing each edge in the bipartite graph and $\bm{J}$ has $N+M$ rows representing both the treatment and trial nodes. 
For a battery attached to nodes $v_j$ and $v_k$, the current is $+1$ at node $v_j$, $-1$ at $v_k$, and 0 at every other treatment node ($v_{l}; l\neq j,k$) and every trial node ($u_i; i=1,\hdots,M$). 
The first $M$ rows of $\bm{J}$ represent trial nodes which are always associated with zero current. 
The remaining $N$ rows represent treatments whose value of current depends on the placement of the battery (in the same way as the unipartite framework). 
Therefore, we can write $\bm{J}=[\bm{0}_{M\times(N-1)} \hspace{3pt} \bm{C}_N']'$ where $\bm{0}_{M\times(N-1)}$ is an $M \times (N-1)$ matrix of zeroes. 
An example of this matrix is shown in Section \ref{app:psoriasis}.

Ohm's law and Kirchoff's law apply in the same way as before leading to the expression
\begin{align}
    \bm{I} = \bm{R}^{-1} \bm{B}_{\top\bot} (\bm{B}_{\top\bot}' \bm{R}^{-1} \bm{B}_{\top\bot})^{+} \bm{J},
\end{align}
which depends on the (oriented) edge-vertex incidence matrix of the \textit{bipartite} graph, $\bm{B}_{\top\bot}$. Matrix $\bm{I}$ has dimensions $K_{\top\bot} \times (N-1)$; each element $I_{jk}^{(il)}$ (row $il$ and column $jk$) is the current flowing through edge $[u_i,v_l]$ when a battery is attached to nodes $v_j$ and $v_k$. The arm-level hat matrix, on the other hand, has dimensions $(N-1) \times K_{\top\bot}$ with rows representing treatment comparisons and columns representing edges. To compare these matrices, we require the transpose of $\bm{I}$,
\begin{align}
    \bm{I}' &= \bm{J}' ((\bm{B}_{\top\bot}' \bm{R}^{-1} \bm{B}_{\top\bot})^{+})' \bm{B}'_{\top\bot} (\bm{R}^{-1})'.
\end{align}
From the definition of a pseudo-inverse it is possible to show (\cite{Stoer:2002}) that $(\bm{A}^{+})'=(\bm{A}')^{+}$ for a general matrix $\bm{A}$. Using this, and the fact that matrices $\bm{R}$ and $\bm{L}=\bm{B}_{\top\bot}' \bm{R}^{-1} \bm{B}_{\top\bot}$ are symmetric ($(\bm{R}^{-1})'=\bm{R}^{-1}$ and $\bm{L}'=\bm{L}$), we find
\begin{align}\label{Eq:supp-Iprime}
    \bm{I'} = \bm{J'}(\bm{B_{\top\bot}'}\bm{R}^{-1}\bm{B_{\top\bot}})^{+}\bm{B_{\top\bot}'}\bm{R}^{-1},
\end{align}
which is the result given in Equation (\ref{eq:Iprime}) in the main paper.

\subsubsection{Random walks and electrical networks}

The random walk on the electrical network can be constructed in the same way as before by defining transition probabilities proportional to the inverse of the resistance in each edge. That is, the probability of hopping from one node to another is proportional to the inverse of the resistance (variance) associated with the edge connecting the two nodes, normalised with respect to the weighted degree of the first node. Since there are no edges between nodes of the same type, the walker can only hop from trial to treatment or from treatment to trial. These probabilities therefore take the same definition as those of the higher order random walk (HoRW) defined in Section \ref{sec:RW-bi} of the main text. 

Without the NMA interpretation of the nodes and edges, the electrical network and random walk defined on this graph have the same properties as before. The differences are: (i) that edges only appear between trial nodes $u_i$ and treatment nodes $v_j$, and (ii) we only consider battery placements between two treatment nodes (since our interest focuses on comparisons between treatments). The analogy can then be written as follows: When a voltage is applied between two (treatment) nodes $a$ and $b$ such that the total current flowing into $a$ and out of $b$ from the exterior is 1, the current induced in edge $[u_i,v_j]$ is equal to the expected net number of times a random walker, starting at $a$ and walking until it reaches $b$, moves along the edge from $u_i$ to $v_j$.

This analogy relates the bipartite edge currents $I_{jk}^{(il)}$ to the movement of a random walker. Conjecture \ref{conj:rw-bi} proposes an equivalent random walk interpretation for the elements of the arm-level hat matrix $H_{\top\bot, il}^{(jk)}$. Therefore, to prove the conjecture it suffices to show that $\bm{I}'$ in Equation (\ref{Eq:supp-Iprime}) is equal to the arm-level hat matrix in Equation (\ref{eq:H-bi}) in the main text.

\newpage
\section{Application}\label{app:psoriasis}
In the following we present various matrices calculated for the psoriasis dataset described in Section \ref{sec:eg} of the main text. In calculating these matrices, we order the treatments and trials alphabetically such that $(v_1,\hdots,v_7)=$ (ETN, IXE\_Q2W, IXE\_Q4W, PBO, SEC\_150, SEC\_300, UST) for treatments and $(u_1,\hdots,u_9)=$ (CLEAR, ERASURE, FEATURE, FIXTURE, IXORA-S, JUNCTURE, UNCOVER-1,  UNCOVER-2, UNCOVER-3) for trials. Treatment ETN was used as the baseline and we chose a common effect (CE) model throughout. We order pairwise combinations of treatments according to the convention $([v_1,v_2], [v_1,v_3], \hdots, [v_1,v_N], [v_2,v_3], [v_2,v_4],\hdots,[v_2,v_N],\allowbreak\hdots,\allowbreak [v_{N-1},v_N])$. We order treatment-trial combinations according to the convention $([u_1,v_2], \hdots, [u_1,v_N], [u_2,v_1],\hdots,\allowbreak[u_2,v_N],\allowbreak\hdots,\allowbreak[u_M,v_1],\allowbreak\hdots, [u_M,v_N])$.

\subsection{Conjecture 1}
\subsubsection{Arm-level hat matrix}
The psoriasis dataset involves $M=9$ trials, $N=7$ treatments and $K_{\top\bot}=28$ arms. 
Each arm is associated with a measure of the log odds of achieving a 75\% improvement on the PASI scale. 
We collect these observations in the vector $\bm{\mu}$. 
Each $n_i-$armed trial is associated with $n_i-1$ relative treatment effects measured as log odds ratios. 
We collect these in the vector $\bm{y}$. 
The vector $\bm{\mu}$ has length 28 and $\bm{y}$ has length $\sum_{i=1}^{M}(n_i-1)=19$.
These two vectors are related via the $19\times28$ matrix $\boldsymbol{C}$ according to Equation (\ref{eq:C}). 
The variances and covariances associated with the log odds ratio measurements are contained in the $19\times19$ matrix $\boldsymbol{\Sigma}$. 
In the CE model the weight matrix is given by its inverse, $\bm{W}=\bm{\Sigma}^{-1}$. 
The design matrix $\boldsymbol{X}$ describes which treatment comparisons appear in each trial and has dimensions $19\times6$. 
All of these vectors and matrices are presented on page \pageref{eq:mu-psor} alongside the resulting trial-level hat matrix, $\boldsymbol{H}$, obtained via Equation (\ref{eq:H-trial}) in the main text, and the arm-level hat matrix $\boldsymbol{H}_{\top\bot}$ obtained via Equation (\ref{eq:H-bi}).

\subsubsection{Matrix of nodal currents}

To check Conjecture 1 for the psoriasis example, we verify that the arm-level hat matrix is the same as the matrix of nodal currents calculated via Equation (\ref{eq:Iprime}) in the main paper. 
This calculation requires the oriented edge-vertex incidence matrix of the bipartite graph, $\boldsymbol{B}_{\top\bot}$, the matrix of edge resistances, $\bm{R}$, and the matrix of edge currents, $\boldsymbol{J}$. 
For the psoriasis network, $\boldsymbol{B}_{\top\bot}$ has $N+M=16$ columns and $K_{\top\bot}=28$ rows. The $28\times28$ matrix of resistances, $\bm{R}$, contains on its diagonal the arm-level variances associated with each log odds measurement in $\bm{\mu}$. The matrix of edge currents has dimensions $6\times16$. Each of these matrices, along with the resulting matrix of nodal currents, $\boldsymbol{I}'$, are shown on page \pageref{eq:Bbi-psor}. 

Comparing each element of $\boldsymbol{I}'$ with the corresponding element of $\bm{H}_{\top\bot}$ gives a maximum difference of $2.44\times10^{-15}$. Therefore, for the psoriasis example, these two matrices are identical within machine precision.

\begin{landscape}
\begin{center}  % Center the rotated equations
%\resizebox{!}{\textwidth}{
\resizebox{\linewidth}{!}{
    %\rotatebox{90}{%
        % Rescale to fit within page height
            $\begin{aligned}\label{eq:mu-psor}
                \bm{\mu} &=     
                \end{aligned}$
        }
        %}
\end{center}
\end{landscape}

\subsection{Conjecture 2}
\subsubsection{Transition matrices for the bipartite random walk}
To define the random walk on the bipartite graph, we require the weighted biadjacency matrix $\bm{B}$ defined in Equation (\ref{eq:Bweight}) in the main text. 
For the psoriasis data,  $\bm{B}$ has dimensions $9 \times 7$. 
Each element $B_{ij}$ is equal to the inverse variance associated with the arm represented by edge $[u_i,v_j]$ in the bipartite graph. 
The downwards transition matrix $\bm{P}^{\downarrow}$ has dimensions $9\times 7$. 
Each row $i$ contains the probabilities of hopping from trial (top) node $u_i$ to each of the treatment (bottom) nodes. 
The upwards transition matrix $\bm{P}^{\uparrow}$, on the other hand, describes transitions from treatments to trials and its dimensions are reversed ($7\times 9$). 
The inner product of the up and down transition matrices yields the $7\times 7$ two-step transition matrix $\bm{P}$. Each element $P_{jk}$ describes the probability of hopping between two treatment nodes, $v_j$ to $v_k$, in two consecutive steps on the bipartite graph. 
We obtain the renormalised two-step transition matrix $\bm{\tilde{P}}$ by setting the diagonal elements equal to zero and renormalising the remaining probabilities in each row. All of these matrices are shown on page \pageref{eq:B-psor}.

\subsubsection{Transition matrix for the unipartite random walk}
The unipartite graph of the psoriasis dataset has $K_\bot=13$ edges and is described by the $13\times 7$ oriented incidence matrix $\bm{B}_\bot$. Performing the aggregate model on the psoriasis data set yields a diagonal $13\times 13$ weight matrix $\bm{W}_\bot$ where each diagonal element contains the aggregate weight associated with that edge. The resulting aggregate hat matrix $\bm{H}_\bot$ has dimensions $6 \times 13$. Each row $j$ of $\bm{H}_\bot$ defines an evidence flow network from the baseline ($v_1:$ ETN) to treatment node $v_j$. The transition matrix of the random walk on the unipartite graph, $\boldsymbol{T}$, has dimensions $7\times7$ and is calculated via Equation (\ref{eq:Tjk}) in the main paper. 
Each row describes the probability of hopping from one treatment node to each of the others. 
All of these matrices are shown on page \pageref{eq:B-psor}.

Conjecture 2 is fulfilled if $\bm{T}$ is equal to the renormalised two-step transition matrix $\bm{\tilde{P}}$. Comparing each element of these matrices yields a maximum absolute difference of $1.33\times 10^{-15}$. Therefore, for the psoriasis example these two matrices are identical within machine precision. 

\begin{landscape}
\begin{center}  % Center the rotated equations
%\resizebox{!}{\textwidth}{
\resizebox{0.87\linewidth}{!}{
    %\rotatebox{90}{%
        % Rescale to fit within page height
            $\begin{aligned}\label{eq:B-psor}
                \bm{B} &=
                \end{aligned}$
        }
        %}
\end{center}
\end{landscape}

\section{Simulation study}\label{app:simulation}
\subsection{Methods}
\subsubsection{Generating random networks of treatments and trials}

Simulating NMA requires two generation procedures: (i) generating the network structure, i.e. which treatments are compared in which trials, and (ii) generating synthetic outcome data in each trial. To date, simulation studies of NMA have focused on the latter, with authors specifying the structure of their network manually before randomly sampling the outcome data from some underlying model. This first step becomes a laborious task for large networks, especially if characteristics of the network need to be varied. Therefore, simulations of NMA have tended to involve small networks (usually $\leq 5$ treatments) with a limited range of geometries. To the best of our knowledge, only one study (\cite{Kanters:2021}) has attempted to automate the generation of treatment-trial structures, but these networks were restricted to two-arm trials. 

To effectively sample networks with multi-arm trials, we require an algorithm that generates graphs with higher-order structures. In the following, we make use of a simple algorithm for random bipartite graphs.

\subsubsection{Random bipartite graphs}\label{app:gen-bi}
In random graph theory, there exist numerous algorithms for the generation of graphs. The simplest is the Erdös-Renyi (ER) model (\cite{ErdosRenyi:1959, ErdosRenyi:1960}) where it is assumed that, for a fixed number of vertices and edges, all possible graphs are equally likely. This model extends naturally to bipartite graphs. Here, for a fixed number of top nodes $N$, bottom nodes $M$, and edges $K_{\top\bot}$, every possible bipartite graph is equally likely. 

The $\texttt{igraph}$ software in $\texttt{R}$ (\cite{igraph:article, igraph:manual})  has a function $\texttt{sample\_bipartite}$ that generates bipartite graphs according to the ER model. We use this to generate random NMA graphs with $N$ treatments, $M$ trials, and  $K_{\top\bot}$ arms. To sample valid NMA networks we impose two additional restrictions: (i) every trial must contain at least two arms, and (ii) the network must be connected. 

Once we have sampled a random graph using $\texttt{sample\_bipartite}$, we ensure condition (i) is fulfilled by sampling additional arms for every trial node with degree $<2$. If a trial node has degree 0, we sample any two treatment nodes and connect them to that trial. If a trial has degree 1, we sample one additional treatment from the bottom nodes that are not already connected to that trial. 

We enforce condition (ii) by using the \texttt{igraph} function \texttt{count\_components} to count the number of connected components of the graph. If this number is greater than 1 then the network is disconnected. To connect it, we randomly sample two components and a bottom node from each, $v_j$ and $v_k$. Next, we sample a trial node $u_i$ from the neighbours of $v_j$ and $v_k$ (i.e., from all the trials that contain one of those treatments). If $u_i$ is connected to $v_j$, we add a new edge connecting it to $v_k$. Otherwise, if $u_i$ is connected to $v_k$, we add a new edge to $v_j$. If there are no trial nodes connected to either $v_j$ or $v_k$, we sample a trial from the full set of top nodes and connect it to both.

\subsubsection{Simulation set up}

In our simulation, we generate 10,000 random networks using the procedure described in Section \ref{app:gen-bi}. At each iteration we sample the number of treatments $N$ from a uniform distribution between 3 and 50, and the number of trials $M$ from a uniform distribution between 2 and 200. Each trial must have at least two arms, so the minimum number of total arms in the network is $2M$. The maximum number of total arms is $NM$, i.e. every trial involves all possible treatments. In reality, the number of arms in a trial tends to be fairly low (e.g, usually $\leq 6$). Therefore, we sample the number of edges (arms) $K_{\top\bot}$ from a half normal distribution bounded between $2M$ and $NM$ with a standard deviation of $(NM-2M)/2$. This ensures that we sample from the full range of possible arms, but gives precedence to lower numbers. Because of the additional restrictions on the network structure (see (i) and (ii) above), the final sampled network may have more arms than originally specified. 

To calculate the matrices required to check the two conjectures, we need the variance associated with each trial arm. We sample these values from a half-normal distribution with a standard deviation $s$. For each network, we sample $s$ uniformly between 0.5 and 2. To check Conjecture 1, we calculate $\bm{H}_{\top\bot}$ and $\bm{I}'$ for each network and calculate the difference between each corresponding element. For Conjecture 2, we calculate and compare each element of $\bm{T}$ and $\bm{\Tilde{P}}$.

\subsection{Results}

Table \ref{tab:sim-results} summarizes the characteristics of the bipartite NMA graphs sampled in the simulation and their unipartite projections. We show the mean, standard deviation (SD), minimum, and maximum values of various graph metrics across the 10,000 random networks. The definitions of these metrics are given in the table footnote. 

\begin{table}[h]
    \centering
    \caption{\small Characteristics of the 10,000 NMA networks sampled in the simulation.  }
    \tabcolsep=5pt
    \begin{tabular}{p{0.45\textwidth}p{0.08\textwidth}p{0.08\textwidth}p{0.08\textwidth}p{0.08\textwidth}}
    \toprule
         & Mean & SD & Minimum & Maximum \\
         \midrule
        Number of treatments (bottom nodes), $N$ & 26.7 & 13.9 & 3 & 50 \\
        Number of trials (top nodes), $M$ & 99.7 & 57.6 & 2 & 200 \\
        Number of edges in the bipartite graph (arms), $K_{\top\bot}$ & 1097.4 & 1173.1 & 5 & 8853 \\
        Number of edges in the unipartite graph, $K_{\bot}$ & 397.8 & 352.2 & 3 & 1225 \\
        Degree of top nodes:\phantom{partite} mean & 11.1 & 8.6 & 2.1 & 49.5\\
        \phantom{Degree of unipartite nodes:} minimum & 6.8 & 7.2 & 2 & 48\\
        \phantom{Degree of unipartite nodes:} maximum & 16.0 & 9.7 & 3 & 50 \\
        Degree of bottom nodes:\phantom{ite} mean & 44.3 & 37.1 & 1.0 & 191.8 \\
        \phantom{Degree of unipartite nodes:} minimum & 36.8 & 35.4 & 1 & 187\\
        \phantom{Degree of unipartite nodes:} maximum & 52.1 & 38.7 & 2 & 195\\
        Degree of unipartite nodes: mean & 23.4 & 13.5 & 2 & 49\\
        \phantom{Degree of unipartite nodes:} minimum & 21.4 & 13.9 & 1 & 49\\
        \phantom{Degree of unipartite nodes:} maximum & 24.8 & 13.5 & 2 &49  \\
        Density of the bipartite graph & 0.45 & 0.23 & 0.05 & 1.00\\
        Density of the unipartite graph & 0.93 & 0.17 & 0.10 & 1.00\\
        Radius of the unipartite graph & 1.14 & 0.39 & 1 & 6\\
        Mean distance of the unipartite graph & 1.08 & 0.24 & 1.00 & 4.80 \\
        \botrule 
    \end{tabular}
    \\
    \noindent\justifying{\small{The density of a graph is the ratio of the number of edges to the number of possible edges. A density of 1 means the network is fully connected. The distance between two vertices is the minimum number of edges between them. The eccentricity of a node is the maximum distance from that node to any other. The radius of a graph is the minimum eccentricity of any node in the graph. In NMA, the degree of a top node (trial) in the bipartite graph is the number of arms in that trial. The degree of a bottom node (treatment) corresponds to the number of trials that treatment is involved in. The degree of a treatment node in the unipartite graph is the number of other treatments that treatment has been directly compared to.}}
    \label{tab:sim-results}
\end{table}

\paragraph{Conjecture 1:} Across all 10,000 networks, the maximum difference between any two corresponding elements of $\bm{H}_{\top\bot}$ and $\bm{I}'$ was $6.03 \times 10^{-11}$. Therefore, Conjecture 1 holds (within machine precision) for all sampled networks.

\paragraph{Conjecture 2:} Across all 10,000 networks, the maximum difference between any two corresponding elements of $\bm{T}$ and $\bm{\Tilde{P}}$ was $1.15\times 10^{-13}$. Therefore, Conjecture 2 holds (within machine precision) for all sampled networks.

\end{appendices}

\end{document}